\documentclass[preprint,12pt]{elsarticle}

\usepackage{amssymb}
\usepackage{float}
\usepackage[a4paper, margin=1in]{geometry}
\usepackage{etoolbox}
\usepackage{soul} 
\usepackage{algorithm}
\usepackage{algpseudocode}
\usepackage{amsmath}
\usepackage{amssymb}
\usepackage{array}
\usepackage{float}
\usepackage{color, colortbl}
\definecolor{Gray}{gray}{0.9}
\usepackage{hyperref}
\usepackage{tikz,xcolor,hyperref}
\usepackage{lscape}
\usepackage{longtable}
\usepackage{multirow}
\newcommand{\orcid}[1]{\href{https://orcid.org/#1}{\includegraphics[width=8pt]{orcid.png}}}
\usepackage{amsmath}
\usepackage{booktabs}
\usepackage{caption}
\usepackage{xurl}

\usepackage{pifont}
\usepackage{tabularx}
\usepackage{array}
\usepackage{makecell}

\usepackage[capitalize]{cleveref}

\newcommand{\cmark}{\textcolor{green!60!black}{\ding{51}}}
\newcommand{\xmark}{\textcolor{red}{\ding{55}}}
\newcommand{\pmark}{\textcolor{orange}{\ensuremath{\approx}}}
\journal{}

\begin{document}

\begin{frontmatter}

%% Title, authors and addresses

%% use the tnoteref command within \title for footnotes;
%% use the tnotetext command for theassociated footnote;
%% use the fnref command within \author or \address for footnotes;
%% use the fntext command for theassociated footnote;
%% use the corref command within \author for corresponding author footnotes;
%% use the cortext command for theassociated footnote;
%% use the ead command for the email address,
%% and the form \ead[url] for the home page:
%% \title{Title\tnoteref{label1}}
%% \tnotetext[label1]{}
%% \author{Name\corref{cor1}\fnref{label2}}
%% \ead{email address}
%% \ead[url]{home page}
%% \fntext[label2]{}
%% \cortext[cor1]{}
%% \affiliation{organization={},
%%             addressline={},
%%             city={},
%%             postcode={},
%%             state={},
%%             country={}}
%% \fntext[label3]{}

\title{DecoyTrace: Toxic Decoys for Active Defense in Decentralized Federated Learning}

%% use optional labels to link authors explicitly to addresses:
%% \author[label1,label2]{}
%% \affiliation[label1]{organization={},
%%             addressline={},
%%             city={},
%%             postcode={},
%%             state={},
%%             country={}}
%%
%% \affiliation[label2]{organization={},
%%             addressline={},
%%             city={},
%%             postcode={},
%%             state={},
%%             country={}}

\author[inst1]{Pedro Beltrán-López}
\ead{pedro.beltranl@um.es}
\affiliation[inst1]{organization={Department of Information and Communication Engineering},%Department and Organization
            addressline={University of Murcia}, 
            city={Murcia},
            postcode={30100},
            country={Spain}}

\author[inst1]{Enrique Tomás Martínez Beltrán}
\ead{enriquetomas@um.es}
\author[inst1]{Pantaleone Nespoli\corref{corauth}}
\ead{pantaleone.nespoli@um.es}
\author[inst1]{Manuel Gil Pérez}
\ead{mgilperez@um.es}
\author[inst1]{Alberto Huertas Celdrán}
\ead{alberto.huertas@um.es}

\cortext[corauth]{Corresponding author.}

\begin{abstract}
%% Text of abstract
Decentralized Federated Learning (DFL) eliminates the central aggregation server, reducing the single point of observation that traditional defenses against attacks rely on. As a result, peer-to-peer networks become exposed to malicious updates containing backdoors or semantic poisoning, since such updates can remain close to benign ones in the parameter space while behaving very differently. This may evade defenses based on passive parameter inspection. However, existing deception-based defenses have mainly been designed for centralized FL and do not jointly address local observation, poisoning propagation, source attribution, and containment in strictly serverless DFL. To address these limitations, this paper presents \textit{DecoyTrace}, a proactive cyber deception-based defense for strictly serverless DFL environments. \textit{DecoyTrace} deploys a mobile \textit{DecoyNode} that generates decoy challenges using chaotic maps, disseminates a dual model (clean vs. decoy) based on neighbor trust, and evaluates them using three-state semantic metrics. Upon confirmation, a distributed protocol isolates the source and performs a model reset or recovery to preserve training progress. Evaluated across sixty configurations on the NEBULA platform (five datasets, three topologies, and four attack/defense scenarios), \textit{DecoyTrace} systematically restores lost utility. The F1-score remains within 0.03 of the baseline on MNIST/FashionMNIST (mitigating drops of up to 0.37), matches or exceeds the baseline on EMNIST and CIFAR-100, and remains between 0.05 and 0.10 below the baseline on CIFAR-10, the most visually complex convolutional scenario evaluated. Furthermore, containment reduces CPU and network usage by up to two-thirds. These results demonstrate the feasibility of unifying deception, identification, and containment in DFL, while also identifying its limitations in complex tasks and multi-attractor threat models.
\end{abstract}
%%%


\end{frontmatter}

%% \linenumbers

%% main text

\section{Introduction}
\label{sec:introduction}

Cybersecurity has become a critical pillar of the digital society due to the increasing adoption of distributed and critical infrastructures, including Internet of Things (IoT) ecosystems, edge computing environments, hyperconnected services, and collaborative systems~\cite{nespoli2019}. At the same time, cyber threats have evolved towards increasingly persistent and adaptive behaviors, forcing defensive mechanisms to operate in environments where adversaries can dynamically modify their strategies in response to the protections the defenders implement~\cite{Abbas2026adaptive}. In this context, Artificial Intelligence (AI) and Machine Learning (ML) have substantially contributed to improving capabilities in cyber threat prevention, detection, and response~\cite {nespoli2026}. However, a significant proportion of existing defensive mechanisms remain static, relying on fixed rules, predefined signatures, invariant thresholds, or passive filtering strategies. Such approaches may become progressively less effective when an adversary can observe, infer, and adapt to the defensive patterns deployed by the system.

Such a limitation has motivated growing interest in dynamic and active defense strategies that deliberately modify the attacker's perception of the environment and induce observable adversarial behavior. In this context, Cyber Deception represents one of the most prominent approaches within this paradigm. It introduces deceptive assets, information, or interactions to increase adversarial uncertainty and encourage attackers to reveal their intentions through controlled engagement~\cite{beltran2026}. Traditionally, Cyber Deception mechanisms have been applied at the network, system, and application layers through techniques such as honeypots, decoys, honeytokens, and Moving Target Defense (MTD), among others. Nevertheless, the same principles are increasingly relevant to emerging distributed learning environments, where the entities participating in the training process themselves constitute part of the attack surface~\cite{Karunamurthy2025FL}.

One particularly relevant domain is Federated Learning (FL), which enables multiple participants to collaboratively train a shared model while keeping their raw data locally. Especially, this paradigm is attractive in scenarios involving privacy, regulatory, organizational, or communication constraints, where centralized collection of training data may be undesirable or infeasible~\cite{kike2025}. However, the distributed nature of FL also introduces important security challenges. Since participating clients directly influence the evolution of the shared model, malicious or compromised participants may manipulate their local training processes and submit poisoned updates~\cite{BELENGUER2025}. Such attacks include model poisoning, targeted label manipulation, and backdoor attacks, among others, where adversarial behavior can be embedded into the collaborative model while preserving apparently legitimate performance on clean inputs.

These challenges become particularly significant in Decentralized Federated Learning (DFL). In fact, DFL removes the central aggregation server, allowing participants to exchange and aggregate model updates directly via peer-to-peer communication~\cite{kike2023}. Eliminating the central coordinator provides several advantages, including improved resilience against single points of failure and greater suitability for highly distributed environments. Nevertheless, this architectural change also removes the trusted global observation point on which many conventional FL defenses implicitly rely~\cite{Rahdari2025survey}. In centralized FL, the aggregation server can inspect updates from multiple clients, maintain global historical information, compare participants, and coordinate mitigation actions. In strict DFL, by contrast, each participant has only a partial and local view of the federation.

Moreover, the security implications of this difference extend beyond the detection of malicious activity. A poisoned model generated by an attacker may be aggregated by an honest neighbor and subsequently propagated to other participants~\cite{fang2024byzantine}. Consequently, a node exhibiting malicious semantic behavior may not necessarily correspond to the original attacker, i.e., it may instead be an honest participant that has unknowingly incorporated and retransmitted poisoned information. In this work, such intermediate participants are referred to as \emph{carriers}. This propagation phenomenon introduces an important distinction among four closely related security problems: detecting the presence of malicious behavior, attributing that behavior to its original source, containing the attacker once identified, and recovering the honest participants' models from any contamination already absorbed before containment took effect. In a decentralized topology, solving the first problem does not necessarily solve the remaining three; i.e., detection does not imply attribution, containment, or recovery.

Existing defenses against malicious participants in FL have predominantly addressed poisoning through robust aggregation, statistical filtering, anomaly detection, or malicious client identification~\cite{Feng2025survey}. Although these mechanisms have proven effective under various assumptions, many rely on a centralized aggregation server or an equivalent global view of the federation. More recently, deception-based approaches have explored using honeypots, decoy models, and controlled backdoors to elicit observable responses from malicious participants~\cite{Zheng2025HoneyFL}. Such mechanisms demonstrate that active probing can reveal information that may remain hidden when exclusively analyzing model parameters~\cite{Sayari2025deception}. However, existing deception-based approaches have mainly been designed for centralized FL environments and therefore do not directly address the combination of local observation, transitive poisoning propagation, source attribution, and distributed containment that inherently arises in serverless DFL.

To address these limitations, this work proposes \textbf{DecoyTrace}~\cite{beltran2026nebula}, a proactive cyber deception-based defense specifically designed for DFL environments without a central aggregation server. Specifically, DecoyTrace introduces an itinerant \textit{DecoyNode} that actively audits neighboring participants by generating dynamic deceptive information, forcing potentially malicious nodes to reveal evidence of malevolent behavior that might otherwise remain hidden under conventional FL-based inspection.

DecoyTrace generates ephemeral bait challenges based on chaotic dynamics and employs a dual-model architecture to avoid contaminating legitimate training. That is, verified neighbors receive a clean model, whereas unverified participants receive a bait model containing a controlled semantic challenge. Consequently, their responses are evaluated using behavioral metrics that aim to determine whether the bait is absorbed, resisted, or replaced by an alternative poisoning objective. However, since poisoned information can propagate through honest participants, semantic evidence alone may reveal an attack without uniquely identifying its source.

To address this attribution problem, DecoyTrace combines semantic verification with a reputation-asymmetry mechanism to distinguish the original attacker from honest carriers. When further investigation is required, the DecoyNode proactively moves through the federation using topology-aware Depth-First Search (DFS) pivoting and distance-guided navigation. Once the attack source is confirmed, DecoyTrace isolates the malicious participant and can trigger a coordinated model recovery mechanism to mitigate residual poisoning effects.

The main contributions of this work can be summarized as follows:

\begin{itemize}

    \item \textbf{An active deception-based architecture for strict DFL}: e introduce DecoyTrace, an open-source framework~\cite{beltran2026nebula}, and the itinerant DecoyNode role, which actively audits a serverless federation through dynamically generated bait challenges while operating exclusively through peer-to-peer interactions.

    \item \textbf{A dual-model and semantic behavioral verification mechanism}: DecoyTrace separates legitimate training from defensive probing through clean and bait models and evaluates received models according to their behavioral response to controlled semantic challenges, rather than relying exclusively on parameter-space inspection.

    \item \textbf{A distributed mechanism for distinguishing attack sources from honest carriers}: We identify and characterize the attribution ambiguity introduced by transitive poisoning propagation in DFL and introduce a reputation-asymmetry mechanism that complements local semantic evidence to distinguish the original attacker from honest participants that inadvertently relay poisoned information.

    \item \textbf{A topology-aware tracing, containment, and recovery strategy}: DecoyTrace combines DFS-based DecoyNode pivoting with distance-guided navigation, localized attacker isolation, and configurable recovery strategies, thereby extending the defensive process beyond malicious-client identification towards actionable mitigation.

    \item \textbf{An extensive experimental validation under heterogeneous DFL conditions}: DecoyTrace is evaluated on the NEBULA platform~\cite{nebula} across five datasets, three network topologies, and four attack/defense scenarios, resulting in $5 \times 3 \times 4 = 60$ experimental configurations. The evaluation analyzes both the predictive utility recovered after poisoning and the associated computational and communication costs under strict serverless operation.

\end{itemize}

The remainder of this article is organized as follows. Section~\ref{sec:sota} reviews the state of the art in malicious-client detection, poisoning mitigation, and cyber deception-based defenses for FL. Section~\ref{sec:methodology} presents the threat model and the complete design of DecoyTrace, including bait generation, dual-model dissemination, semantic verification, distributed attribution, DFS-based tracing, and containment. Section~\ref{sec:experimental-setup} describes the experimental methodology, datasets, network topologies, adversarial configuration, and training setup. Section~\ref{sec:results} presents the quantitative results obtained across the experimental matrix. Section~\ref{sec:discussion} discusses the main findings, their implications with respect to the existing literature, and the limitations of the current approach. Finally, Section~\ref{sec:conclusions} concludes the paper and outlines future research directions.

\section{State of the Art}
\label{sec:sota}

Defending FL against malicious participants has been extensively studied from the perspectives of robust aggregation, statistical filtering, and adversarial client detection. In particular, most of these approaches seek to infer whether a participant is behaving maliciously from the model updates it provides, whereas more recent proposals have moved towards behavioral probing and proactive defenses. In parallel, Cyber Deception techniques have started to introduce honeypots and decoy models into federated environments, among other deceptive techniques. These research directions are closely related to DecoyTrace, but differ substantially in their assumptions regarding coordination, attribution, and response.

Among malicious client detection mechanisms, FLDetector~\cite{ZhangCJG22} proposes detecting malicious clients by observing inconsistencies across multiple rounds of their updates. In this work, the server predicts the expected update of each client using its history and marks anyone who repeatedly deviates as malicious, thereby enabling robust aggregation over the remaining subset. This proposal does not rely solely on a single snapshot per round, but rather on a temporal pattern. Nevertheless, it assumes a central server with the capability to observe and model the history of each client.

FedGT~\cite{Xhemrishi2023} addresses the problem under \textit{secure aggregation} through \textit{group testing}, that is, it organizes clients into overlapping groups, observes aggregates per group, and decodes the presence and identity of malicious ones. Then, it removes them from training, demonstrating effectiveness across various datasets and poisoning attacks. Although it improves the balance between privacy and security, its operation is also server-driven and focuses on identification rather than topological containment in serverless networks.

SafeFL~\cite{Dou2025} shifts the focus from parameter-based comparison towards behavioral verification. Specifically, the server generates synthetic samples and evaluates client models based on their responses, enabling malicious participants to be distinguished without relying solely on distances between model updates. This behavioral perspective is interesting because poisoned models may remain close to benign ones in the parameter space yet behave differently on specific inputs. Nevertheless, SafeFL still relies on a centralized coordinator to construct the auditing data and perform the corresponding evaluation.

Additionally, a second line of research considers more proactive strategies that deliberately modify the information exposed to participants or actively stimulate their models. In this sense, ARMOR~\cite{Zhang2023ARMOR} introduces \textit{differential model distribution}, assigning different models to different clients to reduce the transferability of adversarial attacks generated by Byzantine clients, thereby reinforcing the system's differential robustness. This strategy is oriented toward preventive robustness rather than toward identifying and containing the source. More recently, SpecShield~\cite{hu2026specshield} introduces an active probing mechanism for model-poisoning detection. The server applies calibrated adversarial perturbations using the Fast Gradient Sign Method (FGSM) to elicit characteristic responses from client models, then analyzes these responses in the frequency domain via the Discrete Wavelet Transform (DWT). This strategy allows benign and malicious participants to be distinguished under Non-IID conditions and various poisoning attacks, while avoiding the omission of potential containment mechanisms.

Recent research has also addressed poisoning resilience directly in DFL. In particular, BALANCE~\cite{fang2024byzantine} proposes a Byzantine-robust aggregation mechanism specifically designed for serverless, peer-to-peer DFL. Each participant uses its own local model as a reference for similarity to evaluate received models and filter out potentially malicious updates before aggregation. The approach provides theoretical convergence guarantees under poisoning attacks and demonstrates that Byzantine robustness can be achieved through exclusively local interactions. Nevertheless, BALANCE primarily aims to prevent malicious models from influencing local aggregation and does not address the challenge of tracing the original source of poisoning after malicious information has propagated through honest intermediate participants.

REDFL~\cite{redfl2025} follows a complementary decentralized strategy based on committee consensus. It combines mutual-information-based model validation with participant-contribution assessment to select reliable committee members, which jointly validate model updates, reach consensus, and aggregate the resulting model. This demonstrates an alternative mechanism for achieving poisoning resilience without a permanent central aggregation server. However, similarly to BALANCE, its main objective is robust validation and aggregation rather than actively eliciting adversarial behavior or reconstructing the propagation path of poisoned information to attribute and contain its original source.

Cyber Deception takes the active defense principle one step further by proactively exposing adversaries to controlled deceptive artifacts. In this scenario, HoneyFL~\cite{Zheng2025HoneyFL} introduces honeypot clients that actively attempt to capture backdoor behavior during federated training. The defensive configuration is periodically modified to make the honeypot harder for adaptive adversaries to detect. Similarly, HoneyFED~\cite{MasseBTHSH25} employs a global decoy model to deceive poisoning attackers and capture their attack patterns, inducing the adversary to interact with a model that appears to represent a successful poisoning process. Both approaches demonstrate the potential of Cyber Deception for FL security, but their defensive logic is formulated around centralized federated architectures and, consequently, benefits from a coordinating entity with a global view of the federation.

Taken together, these works reveal a progressive transition from passive inspection towards active interaction with potentially malicious participants. That is, FLDetector and FedGT primarily infer malicious behavior from update consistency or group-level observations, while SafeFL and SpecShield introduce behavioral probing. Moreover, ARMOR proactively modifies model distribution, whereas BALANCE and REDFL demonstrate that poisoning resilience can be achieved in serverless DFL through local filtering and distributed consensus, respectively. Last but not least, HoneyFL and HoneyFED explicitly incorporate deceptive techniques directly into federated training. However, these capabilities remain fragmented across different architectural assumptions and defensive objectives.

In particular, existing active probing and deception-based approaches predominantly rely on a central server or an equivalent global observation point, whereas decentralized approaches such as BALANCE and REDFL primarily focus on filtering, validating, or excluding malicious updates. One can argue that these works leave an important gap in strict DFL environments, where poisoned information may propagate through multiple honest participants before reaching the defensive mechanism. In such settings, detecting suspicious behavior does not necessarily reveal its original source, since an honest participant may have incorporated poisoned parameters and subsequently retransmitted their semantic effects. Consequently, detection, source attribution, and containment must be treated as distinct but interconnected problems.

DecoyTrace addresses this gap by introducing active Cyber Deception directly into a strict serverless DFL environment. Rather than terminating the defensive process after detecting or filtering a suspicious model, DecoyTrace combines dynamically generated bait, semantic behavioral verification, distributed source attribution, topology-aware tracing, containment, and recovery. The proposed itinerant DecoyNode allows the defensive capability to move through the federation following evidence of poisoning, while the reputation-asymmetry mechanism complements semantic observations to distinguish the original attacker from honest \textit{carriers}. Table~\ref{tab:sota_comparison} summarizes the main differences between DecoyTrace and the reviewed approaches.

\begin{table*}[t]
\centering
\caption{Qualitative comparison of representative approaches for malicious client detection, poisoning mitigation, proactive defense, and Cyber Deception in FL/DFL.}
\label{tab:sota_comparison}

\scriptsize
\setlength{\tabcolsep}{2.4pt}
\renewcommand{\arraystretch}{1.12}

\begin{tabularx}{\textwidth}{@{}
>{\raggedright\arraybackslash}p{1.5cm}
>{\centering\arraybackslash}p{0.7cm}
>{\raggedright\arraybackslash}X
>{\raggedright\arraybackslash}p{1.35cm}
>{\centering\arraybackslash}p{0.92cm}
>{\centering\arraybackslash}p{0.92cm}
>{\centering\arraybackslash}p{0.90cm}
>{\centering\arraybackslash}p{0.82cm}
>{\centering\arraybackslash}p{0.95cm}
>{\centering\arraybackslash}p{1.05cm}
@{}}

\toprule

\multirow{2}{*}{\textbf{Work}} &
\multirow{2}{*}{\textbf{Ref.}} &
\multirow{2}{*}{\textbf{Main mechanism}} &
\multirow{2}{*}{\makecell[l]{\textbf{Target}\\\textbf{attack}}} &
\multicolumn{6}{c}{\textbf{Defense capabilities}} \\

\cmidrule(lr){5-10}

& & & &
\makecell{\textbf{Active}\\\textbf{probe}} &
\makecell{\textbf{Decep-}\\\textbf{tion}} &
\makecell{\textbf{Identi-}\\\textbf{cation}} &
\makecell{\phantom{\textbf{Source}}\\\textbf{DFL}} &
\makecell{\textbf{Source}\\\textbf{attr.}} &
\makecell{\textbf{Contain.}\\\textbf{/recovery}} \\

\midrule

FLDetector & \cite{ZhangCJG22} &
Multi-round update consistency &
Poisoning &
\xmark & \xmark & \cmark & \xmark & \xmark & \xmark \\

FedGT & \cite{Xhemrishi2023} &
Group testing and decoding &
Poisoning &
\pmark & \xmark & \cmark & \xmark & \xmark & \xmark \\

SafeFL & \cite{Dou2025} &
Synthetic-data behavioral auditing &
Poisoning &
\cmark & \xmark & \cmark & \xmark & \xmark & \xmark \\

ARMOR & \cite{Zhang2023ARMOR} &
Differential model distribution &
Adv. transfer &
\xmark & \xmark & \xmark & \xmark & \xmark & \xmark \\

SpecShield & \cite{hu2026specshield} &
Adversarial probing and frequency analysis &
Poisoning &
\cmark & \xmark & \cmark & \xmark & \xmark & \xmark \\

\addlinespace[0.2em]

BALANCE & \cite{fang2024byzantine} &
Local similarity-based robust aggregation &
\makecell[l]{Byzantine /\\poisoning} &
\xmark & \xmark & \pmark & \cmark & \xmark & \xmark \\

REDFL & \cite{redfl2025} &
Committee consensus and model validation &
Poisoning &
\xmark & \xmark & \cmark & \cmark & \xmark & \xmark \\

\addlinespace[0.2em]

HoneyFL & \cite{Zheng2025HoneyFL} &
Honeypots (HoneyDoor/HoneyMap) &
Backdoor &
\cmark & \cmark & \cmark & \xmark & \xmark & \pmark \\

HoneyFED & \cite{MasseBTHSH25} &
Global decoy model &
Poisoning &
\cmark & \cmark & \pmark & \xmark & \xmark & \pmark \\

\midrule

\textbf{DecoyTrace} & - &
\textbf{Dynamic decoys, semantic verification, and distributed tracing} &
\makecell[l]{Backdoor /\\Semantic\\poisoning} &
\cmark & \cmark & \cmark & \cmark & \cmark & \cmark \\

\bottomrule
\end{tabularx}

\vspace{0.4em}

\begin{minipage}{\textwidth}
\scriptsize
\textit{Legend:} \cmark\ Yes \quad
\xmark\ No \quad
\pmark\ Partially.
\textit{Source attr.} denotes source attribution beyond the observed intermediate participant.
\textit{Contain./recovery} denotes actions beyond individual-update rejection, such as attacker isolation
or recovery from previously propagated contamination.
\end{minipage}

\end{table*}

Read across its capability columns, Table~\ref{tab:sota_comparison} shows that malicious client identification is already addressed by several existing approaches, while proactive probing and Cyber Deception have also emerged as complementary defensive strategies. In particular, SafeFL and SpecShield actively probe participant behavior, whereas HoneyFL and HoneyFED explicitly incorporate deceptive mechanisms. In parallel, BALANCE and REDFL demonstrate that poisoning resilience can be achieved in serverless DFL through local filtering and distributed consensus. However, these capabilities remain largely separate: active probing and deception-based approaches rely on centralized coordination, whereas decentralized approaches focus primarily on validating or rejecting suspicious updates.

The clearest distinction appears in source attribution and post-detection response. To the best of our knowledge, none of the approaches reviewed in this section explicitly addresses the problem of tracing the original poisoning source once its effects have propagated through honest intermediate participants in strict serverless DFL settings. Likewise, containment and recovery beyond individual update rejection are either absent or only partially addressed in the reviewed literature. Among the approaches compared in Table~\ref{tab:sota_comparison}, DecoyTrace is the only proposal that jointly provides active probing, Cyber Deception, strict serverless DFL operation, explicit source attribution beyond carriers, and coordinated containment/recovery. This combination, rather than any individual capability in isolation, constitutes the main gap highlighted by the comparison.

\section{Methodology and System Design}
\label{sec:methodology}

This section presents the complete design of \textit{DecoyTrace}, a proactive cyber deception-based defense for DFL networks. Unlike reactive approaches, which filter incoming updates using static rules based on model-parameter thresholds, \textit{DecoyTrace} deploys a mobile \textit{DecoyNode} that combines five complementary capabilities, as explained throughout this section.

\subsection{Threat Model}
\label{sec:methodology-threat-model}

The threat model considers a DFL network of \(N\) participants \(\mathcal{P}\), \(|\mathcal{P}| = N\), which collaborate to train a shared model without a coordinator or central server, communicating via a \textit{peer-to-peer} topology, as detailed in Section~\ref{sec:experimental-matrix}. A single attacking participant \(n_a \in \mathcal{P}\) is assumed to exist. Node \(n_a\) has full white-box control over its own local training procedure and can send arbitrary updates to its neighbors in each round. Its goal is to cause the global model to misclassify inputs belonging to a specific class (a targeted label-flipping or backdoor attack), thereby evading detection for as long as possible. No restrictions are imposed on the attacker’s local optimization strategy; that is, it may employ any poisoning technique, local data manipulation, or degree of camouflage in the parameter space. This is because the \textit{DecoyTrace} detection layer is explicitly designed to be agnostic with respect to parametric camouflages, focusing exclusively on evaluating behavior.

It is assumed that the attacker $n_a$ does not have access to the current round’s \textit{decoy} seed \(\sigma_0\) or the trigger pattern \(\Delta\) before receiving a bait model \(\theta_{\text{bait}}\) from the DecoyNode. Furthermore, the attacker cannot distinguish \(\theta_{\text{bait}}\) from \(\theta_{\text{clean}}\) before inspecting it; therefore, the diffusion of the dual model described in Section~\ref{sec:methodology-overview} is not observable a priori by neighbors. Nor is it assumed that the attacker has any external information about which physical node currently serves as the DecoyNode. Following Kerckhoffs’ principle~\cite{petitcolas2025kerckhoffs} for evaluating the worst-case scenario, the attacker is assumed to have full knowledge of the algorithm and the thresholds of \textit{DecoyTrace}. Thus, security relies solely on the secrecy of each round’s chaotic seed, not on the detection logic.

Within this threat model, this work assumes that the attacker $n_a$ does not participate in the network's reputation protocol, i.e., it neither issues nor forwards accusations against other participants. This is adopted here as an explicit assumption of the threat model, rather than as a property derived from rational behavior: although abstaining from reporting plausibly reduces the attacker's exposure, issuing reputation reports would not by itself compromise the attacker's poisoning objective. Consequently, the reputation-asymmetry mechanism of Section~\ref{sec:methodology-attribution} (Eq.~\ref{eq:silent-sink}) is only guaranteed to hold under this non-participation assumption, and Section~\ref{sec:discussion} discusses, as a limitation, the case of an attacker that deviates from it by actively engaging in the reputation protocol.

\subsection{Architectural Overview}
\label{sec:methodology-overview}

\begin{figure}[t]
    \centering
    \includegraphics[width=\linewidth]{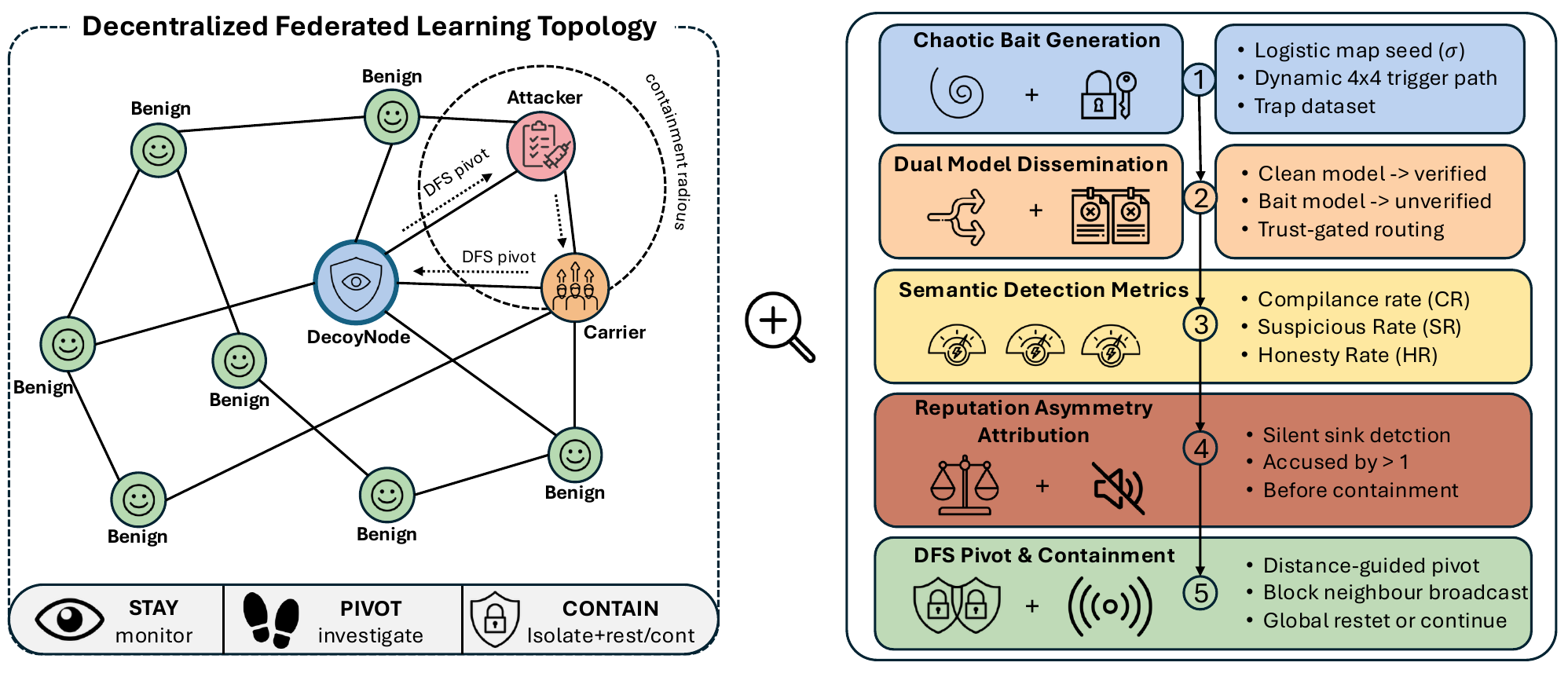}
    \caption{General architecture of \textit{DecoyTrace}. Left: DFL topology where the mobile DecoyNode monitors its neighborhood, pivots via DFS toward suspicious nodes, and contains the threat (STAY, PIVOT, and CONTAIN states). Right: the five functional blocks executed sequentially in each audit round, along with their internal mechanisms.}
    \label{fig:architecture}
\end{figure}

\textit{DecoyTrace} operates as a mobile agent within the DFL topology. At any given time, exactly one node in the network serves as the DecoyNode (the active defender). At the same time, the rest of the participants operate as usual, unaware of which of their neighbors holds that role. This single-active-agent design stems from a deliberate decision to conserve resources: performing bait generation, dual model fine-tuning, and semantic inference simultaneously across all nodes would multiply the computational and network overhead. This is unnecessary, since the primary objective is detection, not universal monitoring. Instead, its itinerant nature allows the defensive capability to follow the trail of evidence across the topology, concentrating the cost exclusively where suspicions exist. Figure~\ref{fig:architecture} illustrates the system’s functional flow as it moves through the network and, on the right, the internal structure of the DecoyNode itself. This is organized into five functional blocks executed sequentially in each audit round: chaotic decoy generation, dual-model propagation, semantic detection metrics, attribution via reputation asymmetry, and, finally, DFS pivoting and containment. Each of these blocks is detailed below, with an explicit justification of its design in light of the failure scenario it mitigates, before being formalized for integration into the full lifecycle in Section~\ref{sec:methodology-lifecycle}.

\subsubsection{Generating Chaotic Baits}
\label{sec:methodologybait}

The robustness of a bait in deep learning depends on two properties: the agent's indistinguishability from a legitimate node and the entropy of its audit pattern. A static verification scheme (e.g., a fixed trigger pattern and a constant target label reused throughout training) would enable an adaptive adversary to apply selective unlearning or gradient-masking techniques. These techniques would be specifically tuned to evade detection by that recurring stimulus, since the challenge would effectively become a known, memorizable constant from the attacker's perspective. To mitigate this threat, \textit{DecoyTrace} generates the decoy using a one-dimensional logistic map~\cite{may1976logisticmap}:

\begin{equation}
    \sigma_{t+1} = r \cdot \sigma_t \cdot (1 - \sigma_t)
    \label{eq:logisticmap}
\end {equation}
\noindent where \(\sigma_t \in (0,1)\) is the internal state of the generator in round \(t\). The endpoints \(0\) and \(1\) are excluded because both are absorbing under Eq.~\ref{eq:logisticmap}, since \(f(0) = f(1) = 0\) for any \(r\), which would collapse the sequence to a trivially predictable fixed point and defeat the unpredictability property this section relies on; moreover, for \(r < 4\) the image of \(f\) on \((0,1)\) never reaches either endpoint, so no interior seed can escape this open interval through the map's own evolution. The value \(r = 3.99\) is adopted from the range \(r \in [3.57, 4)\) established in the chaos-based cryptography literature as the operating region for the logistic map when used as a keyed generator~\cite{mishra2011chaotic}, which places it in its purely chaotic regime. The logistic map exhibits periodic windows for various subintervals of \(r < 4\), so the value \(r = 3.99\) avoids the closest windows while remaining below the domain boundary at \(r = 4\). The system is completely deterministic given the initial seed \(\sigma_0\). This determinism is essential for ensuring the reproducibility of the experiments described in Section~\ref{sec:experimental-setup}. It allows a receiving DecoyNode to regenerate the same sequence of activators following a DFS pivot transfer, as described in Section~\ref{sec:methodology-pivot}. However, the system is computationally unpredictable without access to that seed. The extreme sensitivity to initial conditions, characteristic of chaotic systems, prevents an external observer from predicting \(\sigma_{t+1}\) from a finite sequence of previous outputs without recovering \(\sigma_0\) with full floating-point precision.

From \(\sigma_t\), a label permutation \(\pi_t: \mathcal{Y} \to \mathcal{Y}\) is derived, mapping each original class to a distinct decoy label:
\begin{equation}
    y' = \pi_t(y) = \left(y + 1 + \left\lfloor \sigma_t \cdot (|\mathcal{Y}|-1) \right\rfloor\right) \pmod{|\mathcal{Y}|}
    \label{eq:honeymap}
\end {equation}
\noindent where \(|\mathcal{Y}|\) is the total number of classes, and the offset \(1 + \lfloor \sigma_t \cdot (|\mathcal{Y}|-1) \rfloor\) is restricted by construction to the range \([1, |\mathcal{Y}|-1]\). This guarantees \(y' \neq y\) for every \(y\), since the offset can never be a multiple of \(|\mathcal{Y}|\) and therefore the modulo operation can never return the original label. A naïve offset defined directly over \(\lfloor \sigma_t \cdot |\mathcal{Y}| \rfloor + 1\), by contrast, ranges over \([1, |\mathcal{Y}|]\) and admits the degenerate case in which it equals \(|\mathcal{Y}|\), in which case the modulo operation returns \(y' = y\), silently canceling the decoy signal for that class; the bound used here rules out this edge case entirely. The resulting mapping remains constant throughout the current round, allowing the semantic signal to converge during the local adjustment steps. However, it is regenerated with each new instantiation of the DecoyNode (at each DFS pivot or periodic rotation). Two elements of the bait change from one audit round to the next: the chaotic seed \(\sigma_t\) itself, and consequently the label permutation \(\pi_t\) (Eq.~\ref{eq:honeymap}) and the specific decoy label \(y'\) assigned to each class. Both are regenerated whenever the DecoyNode is instantiated at a new position, either via DFS pivoting (Section~\ref{sec:methodology-pivot}) or periodic rotation, while remaining fixed for the duration of a single round so that the decoy model can converge on that round's mapping.

The previous dynamism is what makes the bait discriminative between honest and malicious participants, as measured through three complementary behavioral indicators formalized later in this section (Section~\ref{sec:methodology-detection}): the honesty rate (\(HR\)), the fraction of trigger-shown samples for which a neighbor's model still predicts the original, unperturbed class; the compliance rate (\(CR\)), the fraction for which it instead predicts the decoy label injected by the current permutation; and the suspicion rate (\(SR\)), the fraction for which it predicts neither, redirecting toward some other, attacker-controlled target class. An honest node has no mechanism to anticipate or memorize the current round's permutation; its response to \(\theta_{\text{bait}}\) is governed solely by its own honest training objective, so it is expected to resist the trigger and correctly classify the original class (high \(HR\), low \(CR\) and \(SR\)) regardless of which specific permutation is in effect that round. A malicious node, by contrast, is expected to either absorb the decoy label (high \(CR\)) or redirect its prediction toward its own poisoning target (high \(SR\)), since its objective is unrelated to correctly resisting whatever trigger happens to be presented. If the bait were static, an adaptive attacker could learn to recognize the fixed trigger-label pair through selective unlearning or gradient-masking, as discussed at the beginning of this section, and thereby respond to it exactly as an honest node would, collapsing the discriminative gap between the two behavioral profiles. Because the mapping instead changes unpredictably between audits and cannot be anticipated without recovering \(\sigma_0\) to full floating-point precision, an attacker cannot pre-condition its model against the specific pattern that will be used in a given round, so its response continues to reflect its underlying malicious objective rather than a learned evasion of a known, static signal.

A high-entropy additive trigger \(\Delta\) is superimposed on the samples from the benign dataset \(\mathcal{D}_{\text{benign}}\): a \(4 \times 4\) pixel white patch inserted in the lower-right corner of each image. This design serves two purposes. First, the patch is spatially located in a region that contributes little discriminative signal to the main classification task in the evaluated datasets, as described in Section~\ref{sec:experimental-matrix}. Therefore, the model can learn to associate the decoy with the trigger without altering its normal decision boundary elsewhere in the image, thereby preserving its baseline performance on clean inputs. Second, its high-contrast intensity is designed to survive standard per-channel normalization. This distinguishes it from low-amplitude additive noise activators, which are typically absorbed or substantially attenuated by normalization. The trap dataset $\mathcal{D}^{(t)}_{\text{trap}}$ with the resulting set for round \(t\) is defined as:
\begin{equation}
    \mathcal{D}^{(t)}_{\text{trap}} = \left\{ (x \oplus \Delta,\ y') \mid (x,y) \in \mathcal{D}_{\text{benign}} \right\}
    \label{eq:trapdataset}
\end{equation}

\noindent where \(x\) and \(y\) are the image and original label of a sample from the benign set \(\mathcal{D}_{\text{benign}}\); \(\Delta\) is the additive activator; and \(y' \) is the decoy label assigned by the mapping in Eq.~\ref{eq:honeymap}. Each sample is stamped with an injection probability \(p_{\text{inj}} = 0.5\). This value is intended to strike a deliberate balance rather than being an arbitrary standard. A value that is too low would result in a weak, inconsistent backdoor that would fail to reliably separate compliant neighbors from resistant ones. Conversely, a value close to 1 would degrade the model's performance on clean tasks, making it an outlier that an adaptive adversary could easily identify. The value \(p_{\text{inj}} = 0.5\) balances the prominence of the trigger and the plausibility of the decoy model. When trained on \(\mathcal{D}^{(t)}_{\text{trap}}\), the decoy model learns a controlled mapping \(\mathcal{C}_t : (x \oplus \Delta) \to y'\). This serves as an ephemeral validation token, unique to each audit round and unrecoverable by an adversary who has not observed the model used in that round.

\subsubsection{Dual Model Diffusion}
\label{sec:methodologydual}

To avoid ``friendly fire'' (i.e., the unintended contamination of honest nodes by gradients derived from the backdoor label of the decoy \(y'\)), \textit{DecoyTrace} maintains two independent parameter vectors. This label is, by design, semantically incorrect for the original input. The first vector, \(\theta_{\text{clean}}\), is trained exclusively on \(\mathcal{D}_{\text{benign}}\) and is reserved for neighbors whose trustworthiness has already been verified. The second, \(\theta_{\text{bait}}\), is a branch of the global model that has been fine-tuned on \(\mathcal{D}^{(t)}_{\text{trap}}\), as defined in Eq.~\ref{eq:trapdataset}, and is reserved for neighbors whose trust status has not yet been established, or whose behavior gives cause for suspicion.

Maintaining a single shared model and testing all neighbors with the same bait-laden update was considered but rejected at the beginning of the design process. This is precisely because it would propagate the backdoor label to honest, already-verified neighbors through standard FedAvg aggregation~\cite{mcmahan2017fedavg}. Such propagation would degrade the model’s utility without providing any forensic benefit. FedAvg is used here, rather than a distance-based robust aggregator such as Krum~\cite{blanchard2017krum} or a coordinate-wise trimmed aggregator, precisely because \textit{DecoyTrace} targets the regime in which robust aggregation by parameter-space distance is insufficient: a semantically poisoned update can remain close to the honest update distribution in parameter space while behaving arbitrarily differently on trigger inputs, as argued in Section~\ref{sec:methodology-detection}, so pairing the evaluation with a distance-based robust aggregator would pre-filter part of the very attack surface this defense is designed to address and confound attribution of the observed recovery. The division into two models isolates this cost, restricting it solely to the subset of the neighborhood that has not yet earned trust. The selective transmission function \(\Psi_{\text{tx}} : \mathcal{N} \to \Theta \cup \{\emptyset\}\) determines the payload sent to each neighbor \(n_j\) based on its trust level \(\mathcal{T}(n_j)\), where \(\Theta\) is the space of model parameter vectors and the adjoined element \(\emptyset\) is a sentinel denoting that no payload is transmitted, not a set-theoretic empty set within \(\Theta\) itself:
\begin{equation}
    \Psi_{\text{tx}}(n_j) =
    \begin{cases}
        \theta_{\text{clean}} & \text{if } \mathcal{T}(n_j) \in \{\text{BENIGN}\} \\
        \theta_{\text{bait}} & \text{if } \mathcal{T}(n_j) \in \{\text{TESTING}\} \\
        \emptyset & \text{if } \mathcal{T}(n_j) = \text{MALICIOUS}
    \end{cases}
    \label{eq:dualmodel}
\end {equation}
\noindent where \(n_j\) is the target neighbor, \(\mathcal{T}(n_j)\) is its current confidence level, and \(\Psi_{\text{tx}}(n_j)\) is the payload that the DecoyNode transmits to it. This payload corresponds to the clean model \(\theta_{\text{clean}}\) for a \texttt{BENIGN} neighbor, the bait model \(\theta_{\text{bait}}\) for a \texttt{TESTING} neighbor, and no payload, denoted as \(\emptyset\), for a neighbor already confirmed as \texttt{MALICIOUS}.

The third branch, which involves withholding any payload once a neighbor is confirmed as \texttt{MALICIOUS}, constitutes a deliberate defensive measure rather than a mere administrative artifact. This action eliminates both the incentive and the opportunity for the confirmed attacker to continue collecting fresh updates from the bait model. Otherwise, the attacker could use these updates to create digital fingerprints and adapt to the current round’s trigger. This design makes DecoyNode a dual-purpose entity. On the one hand, it honestly joins with verified peers. On the other hand, it deploys an active trap exclusively against unverified neighbors, forcing them to reveal their behavior through their response to \(\theta_{\text{bait}}\). All of this is achieved without penalizing global convergence among nodes that have already demonstrated benign behavior.

\subsubsection{Semantic Detection Metrics}
\label{sec:methodology-detection}

Parameter-distance metrics, such as Euclidean distance, cosine similarity between weight vectors, and related statistical outlier-detection schemes over model updates, are insufficient against semantic poisoning attacks. An adversary can preserve close parametric proximity to the honest update distribution while selectively altering the model's behavior only in response to specific triggers, so the poisoned and clean models can be numerically close in parameter space yet behaviorally divergent on the inputs that matter for the attack. \textit{DecoyTrace}, therefore, replaces parameter inspection with behavioral verification, in which a received model \(\theta_{\text{rx}}\), the payload obtained from a neighbor \(n_j\) in the current round, of a kind not yet known to the DecoyNode, since it may be \(\theta_{\text{clean}}\), a neighbor's own bait-laden model, or an arbitrarily poisoned model, is evaluated on the trap validation set \(\mathcal{V}_{\text{trap}}\), a single batch drawn from the DecoyNode's own local validation split and stamped with the trigger \(\Delta\) exactly as in Eq.~\ref{eq:trapdataset}, from which three tri-state classification metrics are derived. Each metric below is a probability estimated as the empirical frequency of the corresponding event over \(\mathcal{V}_{\text{trap}}\), and therefore takes values in \([0,1]\).

The compliance rate, or absorption metric \(CR\), measures whether the neighbor absorbed the controlled backdoor \(\mathcal{C}_t\), that is, whether it predicts the decoy label \(y'\) when presented with the trigger \(\Delta\):
\begin{equation}
    CR(\theta_{\text{rx}}) = P\left(\hat{y} = y' \mid x \oplus \Delta;\ \theta_{\text{rx}}\right) \in [0,1]
    \label{eq:cr}
\end{equation}

Complementing it, the suspicion rate, or malicious metric \(SR\), captures the probability that the model redirects its prediction toward an attack target class \(k\), distinct from both the original label and the decoy label:
\begin{equation}
    SR(\theta_{\text{rx}}) = \max_{k \in \mathcal{Y} \setminus \{y, y'\}} P\left(\hat{y} = k \mid x \oplus \Delta;\ \theta_{\text{rx}}\right)
    \label{eq:sr}
\end{equation}
\noindent Taking the maximum over \(k\), rather than a fixed pre-declared target class, is intentional, because at deployment time the DecoyNode has no prior knowledge of which class the attacker is targeting, so committing to a single fixed \(k\) would make detection trivially evadable by any attacker not targeting that particular class. The maximum formulation instead detects concentration toward any class outside the pair \(\{y, y'\}\), which is what an untargeted or arbitrarily-targeted label-flipping attack produces regardless of its specific target.

The third metric, the honesty rate, or resistance metric \(HR\), evaluates whether the model ignores the trigger altogether and correctly classifies the original class:
\begin{equation}
    HR(\theta_{\text{rx}}) = P\left(\hat{y} = y \mid x \oplus \Delta;\ \theta_{\text{rx}}\right)
    \label{eq:hr}
\end{equation}
\noindent \(CR\), \(SR\), and \(HR\) are computed over overlapping but distinct prediction events, so they are not required to sum to 1, since a given model's per-sample predictions are partitioned across a potentially large set of classes \(\mathcal{Y} \setminus \{y,y'\}\) in the case of \(SR\)'s argmax, and mass can therefore spread across several \(k\) without any single one dominating. All three are reported because each captures a functionally distinct failure or success mode: absorbing the decoy is not the same event as drifting toward an attacker-controlled class, and neither is the same as correctly resisting the trigger.

From these three rates, the system defines a sustained local attack signature, combining an absolute suspicion threshold \(\tau_{\text{susp}} = 0.40\) on \(SR\) with residual-honesty and residual-absorption thresholds, \(\varepsilon = 0.02\) and \(\tau_{\text{low-honest}} = 0.35\), evaluated consecutively over \(\rho = 4\) rounds, the smallest streak length for which zero false positives from aggregation noise alone were observed across the validation sweep, while keeping the additional detection delay it introduces, at most \(\rho - 1\) extra rounds beyond the first true hit, acceptable relative to the total training horizon of \(T = 100\) rounds used in the experiments, described in Section~\ref{sec:experimental-federation}:
\begin{equation}
    \text{semantic\_hit}_t = \mathbb{1}\left[(SR_t \geq \tau_{\text{susp}}) \ \land\ (CR_t < \varepsilon) \ \land\ (HR_t < \tau_{\text{low-honest}})\right]
    \label{eq:semantic-hit}
\end{equation}
\begin{equation}
    \text{local\_streak} = \max\{k : \text{semantic\_hit}_{t-k+1} = \cdots = \text{semantic\_hit}_t = 1\}
    \label{eq:local-streak}
\end{equation}
\noindent where \(t\) indexes global training rounds of the federation, the same index used for \(T = 100\) in Section~\ref{sec:experimental-federation} and for the chaotic state \(\sigma_t\) in Section~\ref{sec:methodologybait}; \(\text{semantic\_hit}_t\) is a binary indicator, denoted \(\mathbb{1}[\cdot]\), equal to 1 when round \(t\) jointly satisfies all three thresholds, and 0 otherwise; \(SR_t\), \(CR_t\), and \(HR_t\) are the suspicion, compliance, and honesty rates of Eqs.~\ref{eq:cr}--\ref{eq:hr} evaluated at round \(t\); \(\tau_{\text{susp}}\), \(\varepsilon\), and \(\tau_{\text{low-honest}}\) are the suspicion, residual-absorption, and residual-honesty thresholds; and \(\text{local\_streak}\) is the length \(k\) of the longest run of consecutive rounds, ending at the current round \(t\), for which \(\text{semantic\_hit}\) held continuously. The three-way conjunction in Eq.~\ref{eq:semantic-hit}, rather than a threshold on \(SR\) alone, rejects a specific false-positive mode observed during development, in which a node undergoing ordinary training noise or class imbalance can occasionally spike on \(SR\) for isolated batches without exhibiting the co-occurring pattern of low decoy absorption and low honest resistance that characterizes an actual attack response. Requiring all three conditions simultaneously rules out that noise source. The thresholds \(\tau_{\text{susp}} = 0.40\), \(\varepsilon = 0.02\), and \(\tau_{\text{low-honest}} = 0.35\) were all calibrated through iterative manual tuning across repeated validation runs during development, rather than a single formal sweep; \(\varepsilon\) in particular was lowered from an initially wider value after repeated runs on ring-shaped sparse topologies showed that a looser residual-absorption threshold let benign, merely diluted compliance signals cross it and trigger false positives, which motivated tightening it to its current value.

The requirement of temporal sustainment, meaning \(\texttt{local\_streak} \geq \rho\), is critical and constitutes the single most important robustness property of this subsection. FedAvg aggregation introduces dilution artifacts in topologies with wide neighborhoods, where an honest node can exhibit isolated spikes in absorption or suspicion rate without this constituting genuine evidence of sustained malicious behavior, since a single round's aggregation can transiently pull a node's response distribution in either direction purely as a numerical artifact of averaging multiple independent updates. Requiring consistency across \(\rho\) consecutive rounds rejects this class of false positives by construction, since it is statistically implausible for uncorrelated aggregation noise to reproduce the same three-way threshold crossing four rounds in a row, while still preserving low detection latency.

\subsubsection{Reputation Asymmetry Attribution}
\label{sec:methodology-attribution}

A systematic forensic analysis of how the semantic signature propagates in dense topologies revealed that the \(CR\), \(SR\), and \(HR\) metrics defined in Section~\ref{sec:methodology-detection}, while effective at detecting the presence of an attack in a neighborhood, are not sufficient to attribute its origin. In a topology using FedAvg aggregation, an honest node \(n_h\) that averages the poisoned gradient of a malicious neighbor \(n_a\) inherits a fraction of its semantic fingerprint, and empirical observation shows that the suspicion rate of honest nodes adjacent to the attacker can match or exceed that of the attacker itself, reaching \(SR_{n_h} = 0.93\) versus \(SR_{n_a} = 0.90\) in the same experimental run. More strikingly, the compliance signal \(CR\), intuitively the most direct indicator of resistance to the bait, is empirically inverted in some configurations, since the attacker, by aggregating the legitimate models of its honest neighbors as part of its own local aggregation step, occasionally leaks a small amount of absorbed bait, \(CR > 0\), whereas a clean honest relay can remain at exactly \(CR = 0\) throughout the entire run. Consequently, no threshold defined purely over \(\{CR, SR, HR\}\) reliably separates the source of the poisoning from its unwitting carriers, regardless of how that threshold is tuned. This is a structural property of the aggregation dynamics rather than a hyperparameter-tuning failure, confirmed by exhaustively sweeping every threshold combination constructible from local metrics alone, including absolute rate thresholds, spatial poison-intensity gradients across the topology, bait-response slope over time, and extreme sustained-rate thresholds up to SR$\geq$0.98, before concluding that instantaneous attribution highly benefits from an optional orthogonal signal to optimize the tracing process.

Underlying the robust discriminator is the observation that the source and the carriers differ systematically in their behavior within the network's reputation protocol, rather than in the semantic fingerprint of their models. By construction of the threat model, described in Section~\ref{sec:methodology-threat-model}, an attacking node does not forward or participate in the network's reputation feedback protocol, since reporting on itself or on other nodes would be directly counter to its objective of remaining unnoticed, whereas an honest relay continues to actively report on its own neighbors as part of the system's normal operation, independent of whether it happens to be unknowingly forwarding poisoned gradients. This asymmetry is not an incidental correlation but a direct, unavoidable consequence of the attacker's own incentive structure, which makes it a structurally robust signal rather than a statistically robust but defeatable one. Formally, for each node \(n_j\in\mathcal{P}\), the asymmetry triplet is defined as:
\begin{align}
    \text{accused\_by}(n_j) &= \left| \{ n_i \in \mathcal{P} \setminus \{n_j\} : n_i \to n_j \in \mathcal{A} \} \right| \label{eq:accused-by}\\
    \text{accuses}(n_j) &= \left| \{ n_k \in \mathcal{P} : n_j \to n_k \in \mathcal{A} \} \right| \label{eq:accuses}\\
    \text{active\_reporter}(n_j) &= \mathbb{1}\left[ n_j \text{ reported within the last } w \text{ rounds} \right] \in \{0,1\} \label{eq:active-reporter}
\end{align}
\noindent where \(\text{accused\_by}(n_j)\) counts the distinct neighbors that have accused \(n_j\); \(\text{accuses}(n_j)\) counts the distinct neighbors that \(n_j\) itself has accused; \(\text{active\_reporter}(n_j)\) is a binary indicator, denoted \(\mathbb{1}[\cdot]\), equal to 1 if \(n_j\) has issued at least one accusation within the freshness window, and 0 otherwise; \(\mathcal{A} \subseteq \mathcal{P} \times \mathcal{P}\) is the set of directed accusation edges observed in the distributed reputation graph, where \(n_i \to n_j \in \mathcal{A}\) denotes that \(n_i\) has accused \(n_j\); and \(w\) is the freshness window itself, chosen to be large enough to tolerate the asynchronous, gossip-based propagation delay of reputation updates across the topology, but short enough to exclude a node's reporting history from before it may have changed roles. A node is classified as a silent sink, the behavioral profile of the attack source, if and only if:

\noindent where \(\text{accused\_by}(n_j)\) counts the distinct neighbors that have accused \(n_j\); \(\text{accuses}(n_j)\) counts the distinct neighbors that \(n_j\) itself has accused; \(\text{active\_reporter}(n_j)\) is a binary indicator, denoted \(\mathbb{1}[\cdot]\), equal to 1 if \(n_j\) has issued at least one accusation within the freshness window, and 0 otherwise; \(\mathcal{A}\) is the set of accusations observed in the distributed reputation graph; and \(w\) is the freshness window itself, chosen to be large enough to tolerate the asynchronous, gossip-based propagation delay of reputation updates across the topology, but short enough to exclude a node's reporting history from before it may have changed roles. A node is classified as a silent sink, the behavioral profile of the attack source, if and only if:
\begin{equation}
    \text{is\_silent\_sink}(n_j) = \mathbb{1}\left[(\text{accuses}(n_j) = 0) \ \land\ \lnot\,\text{active\_reporter}(n_j) \ \land\ (\text{accused\_by}(n_j) \geq 1)\right]
    \label{eq:silent-sink}
\end{equation}

\noindent where \(\text{is\_silent\_sink}(n_j)\) is 1 exactly when node \(n_j\) satisfies all three conditions jointly: it has issued zero accusations of its own; it is not currently an active reporter; and it has been accused by at least one neighbor. Unlike the semantic metrics, this condition is infeasible for an honest carrier to satisfy, as an explicit adversarial-robustness argument rather than merely an empirical observation. A relay that unknowingly forwards poison continues reporting on its own neighborhood as part of its normal, unmodified operation as an honest participant, and therefore necessarily exhibits \(\text{accuses}(n_j) > 0\), failing condition~\eqref{eq:silent-sink} by construction. Conversely, the only way for the true attacker to defeat this discriminator would be to begin actively participating in the reputation protocol, that is, to start behaving indistinguishably from an honest reporter, which would require it either to report truthfully, undermining nothing about its own detection since Eq.~\ref{eq:silent-sink} only checks whether it reports at all and not what it reports, or to report dishonestly, which surfaces as a separate, independently detectable anomaly in the reputation graph, namely a node accusing peers it has no semantic evidence against. That anomaly is outside the scope of this mechanism, but it does not weaken it, since either behavior removes the attacker's silence, which is the only property this gate relies on.

When this optimization is enabled, combining both signals conjunctively produces an accelerated final conviction decision, referred to as the hybrid conviction gate. In this mode, the local signature acts as a fast, necessary trigger and the reputation asymmetry acts as a rapid attribution arbiter to shortcut the standard DFS traceback:

\begin{equation}
    \text{CONVICT}(n_j) = \underbrace{(\text{local\_streak}(n_j) \geq \rho)}_{\text{presence (local)}} \ \land\ \underbrace{\text{is\_silent\_sink}(n_j)}_{\text{attribution (reputation)}}
    \label{eq:hybrid-gate}
\end{equation}
\noindent where \(\text{CONVICT}(n_j)\) is the binary conviction decision for neighbor \(n_j\); the first term, labeled presence, is 1 when the local streak of Eq.~\ref{eq:local-streak} has reached the sustainment threshold \(\rho\); the second term, labeled attribution, is 1 when \(n_j\) satisfies the silent-sink condition of Eq.~\ref{eq:silent-sink}; and \(\land\) denotes logical conjunction, so \(n_j\) is convicted only when both terms hold simultaneously. This design resolves two limitations observed in the reference implementation, each corresponding to a concrete failure mode encountered during development. First, it avoids convicting honest nodes with an elevated semantic signature, since such nodes actively participate in the reputation protocol and therefore never satisfy Eq.~\eqref{eq:silent-sink}, no matter how extreme their local \(SR\) becomes, which directly rules out the false-positive regime observed when using local metrics alone. Second, it avoids the detection latency introduced by requiring a quorum of independent corroborating accusers, by instead requiring corroboration from only a single independent reporter, \(\text{accused\_by} \geq 1\). This second point is critical in sparse topologies, where the attacker may have only one honest neighbor capable of accusing it. A quorum-based requirement, \(\text{accused\_by} \geq 2\), tested and rejected during development, would leave the attacker permanently unconvictable precisely in the low-connectivity regime where topological containment matters most, since the attacker would never accumulate a second independent accuser. For completeness, when using this optimization, Eq.~\ref{eq:hybrid-gate} acts as a conjunction rather than a disjunction because it relies on the synergy between both signals to safely shortcut the slower DFS pivoting process. The reputation signal alone, without the local streak requirement, would convict any node that happens to have low reputation-protocol activity for reasons unrelated to an attack, for instance, a node that recently joined or one experiencing a transient communication fault, so the local trigger keeps the reputation check from being invoked at all except when the network already has strong local evidence of active poisoning under way.

\subsubsection{DFS Pivot and Containment}
\label{sec:methodology-pivot}

When an audited neighbor exhibits a sustained semantic signature but does not satisfy the silent-sink criterion of Eq.~\ref{eq:silent-sink}, \textit{DecoyTrace} classifies it as a carrier, an intermediate node that relays the poison without being its origin. In this case, the system transfers its complete internal state, \((\sigma_t, \mathcal{H}, \theta_{\text{bait}})\), where \(\mathcal{H} \subseteq \mathcal{P}\) is the visited-node history, the set of nodes the itinerant DecoyNode has already occupied during the current depth-first search, initialized to the empty set at the start of a trace and extended by one element at each pivot, as formalized in Section~\ref{sec:methodology-lifecycle}, to the carrier node, which temporarily assumes the DecoyNode role and repeats the audit over its own neighborhood in the following round, generating a distributed depth-first search that walks the infection chain backward, node by node, toward its source. Transferring the full chaotic state \(\sigma_t\), rather than re-seeding a fresh generator at the new DecoyNode, is intentional, since it preserves round-to-round bait continuity for the neighbors that were already under evaluation from the previous position, avoiding the wasted audit rounds that a full generator reset would introduce at every pivot.

The left panel of Figure~\ref{fig:architecture} depicts this distance-guided pivoting alongside the resulting containment radius around the confirmed source. Sparse-connectivity topologies pose a distinct challenge: the attacking node may not be reachable within a small number of hops from the DecoyNode's current position, so relying purely on greedy carrier-to-carrier pivoting, meaning only moving to a directly audited neighbor, can in principle stall if no direct neighbor currently exhibits a sustained local signature even though a silent sink exists elsewhere in the topology. To guarantee search convergence without depending on a strictly local view of the reputation graph, which can legitimately differ between adjacent nodes due to the asynchronous, gossip-based propagation of reputation updates, \textit{DecoyTrace} incorporates a distance-guided navigation strategy. When the accusation graph identifies a silent-sink candidate that is not a direct neighbor, the system computes the topological distance via breadth-first search from its current position to every valid candidate and prioritizes the closest one as the navigation target, rather than the one with the highest accusation count:
\begin{equation}
    n^{*} = \operatorname*{arg\,min}_{n_j \,:\, \text{is\_silent\_sink}(n_j)} \Big( d_{\text{BFS}}(n_{\text{actual}}, n_j),\ -\text{accused\_by}(n_j) \Big)
    \label{eq:nav-target}
\end{equation}
\noindent where \(n^{*}\) is the selected navigation target; the minimization ranges over every node \(n_j\) currently satisfying the silent-sink condition of Eq.~\ref{eq:silent-sink}, that is, every candidate the accusation graph identifies as a plausible attack source; \(d_{\text{BFS}}(n_{\text{actual}}, n_j)\) is the breadth-first-search hop distance between the DecoyNode's current position \(n_{\text{actual}}\) and candidate \(n_j\); and \(\text{accused\_by}(n_j)\), negated so that a higher accusation count ranks more favorably, is the corroboration count of Eq.~\ref{eq:accused-by}. The ordered pair is compared lexicographically, with distance first and accusation count as a tie-breaker. Distance is prioritized over accusation count deliberately, even though a higher accusation count might intuitively seem like stronger evidence, because a distant, heavily accused candidate is, from the current DecoyNode's position, no more actionable in the near term than a closer, less-corroborated one. Prioritizing distance minimizes the number of pivot hops, and therefore the wall-clock rounds, required before the DecoyNode can begin direct, first-hand semantic verification of the candidate, which is a strictly stronger form of evidence than any reputation-graph inference.

Once the target \(n^{*}\) is fixed, the system commits to that navigation target for a budget of \(H = 5\) real hops, measured by the growth of the visited-node history rather than by elapsed rounds or by local reputation re-evaluations, before reconsidering the target. This specific measurement choice was forced by a concrete bug class encountered and iteratively resolved during development. Early designs that re-evaluated the navigation target on every call, rather than committing for a fixed hop budget, were vulnerable to a two-node oscillation, because reputation-graph visibility can differ between two adjacent nodes at the same wall-clock time, due to gossip propagation lag, causing each of the two nodes to alternately appear as the more attractive target to the other and inducing an infinite back-and-forth cycle instead of forward progress. A grace-counter design, allowing a fixed number of re-evaluation strikes before forcing a hop, was also tried and rejected, because the counter was consumed by the evaluation call cadence, which runs faster than the actual pivot cadence, itself gated by a separate per-node grace-period timer, so the counter would exhaust itself before a single real hop had actually occurred, reproducing the same oscillation. Measuring the commitment budget in real, physically executed hops, rather than in calls, timer ticks, or rounds, was the only formulation that eliminated the oscillation in validation runs, converging reliably even in the sparsest topologies tested, an Erd\H{o}s R\'enyi edge probability of \(edge=0.2\), described in Section~\ref{sec:experimental-matrix}.

Once the hybrid gate of Eq.~\eqref{eq:hybrid-gate} confirms the source node \(n_{\text{adv}}\), \textit{DecoyTrace} executes a three-phase containment protocol. The first phase is isolation. The direct neighbors of \(n_{\text{adv}}\) are identified using the globally known topology, and a priority direct connection is established with each of them to guarantee propagation reach for the following control message, independent of the topology's normal gossip-forwarding latency. A \texttt{BLOCK\_NEIGHBOR} control message is then broadcast to those neighbors, which update their access-control list to reject any future updates originating from \(n_{\text{adv}}\). This design targets only the direct neighborhood of the confirmed attacker rather than reconfiguring the topology globally, so as to preserve connectivity across the rest of the network. A global topology reconfiguration would risk partitioning honest regions of the graph that have no direct exposure to the attacker, an unnecessary and potentially destabilizing cost relative to the localized blocking rule.

The second phase is model sanitization. Blocking alone does not remove the poisoned gradients already absorbed into the network's models via FedAvg aggregation in prior rounds, since the attacker's contribution persists in the aggregate weights of every honest node that aggregated with it, directly or transitively, before conviction. \textit{DecoyTrace} supports two configurable strategies to address this residual contamination. The first strategy, coordinated global reset, restores all honest nodes' weights to a synchronized initial state, referred to as \textit{SYNC\_SEED}, at a shared target round, fully purging residual contamination at the cost of discarding all training progress accumulated up to that point. The second strategy, continuation without reset, simply lets the network continue training after the attacker's isolation, relying on the now-uncontaminated aggregation stream to dilute and gradually overwrite the residual poisoned contribution over subsequent rounds. Neither strategy unconditionally dominates the other: a global reset guarantees complete removal of the poisoned contribution at the cost of rolling back training progress, whereas continuation without reset preserves progress but leaves a residual, decaying bias for a short span of rounds after conviction, typically a handful of rounds given that the attacker is confirmed and blocked early in the \(T=100\)-round horizon across the evaluated configurations, well before the point of steady-state convergence in most cases. The quantitative trade-off between the two, across all evaluated datasets and topologies, is presented in Section~\ref{sec:results}.

The third phase is single-shot confirmation. The reset order is issued exactly once per convicted attacker, via a shared target round propagated through gossip, guaranteeing synchronized execution across all nodes without redundant retries that would otherwise unnecessarily degrade convergence. This one-shot guarantee is enforced independently at both the broadcasting node, which records that it has already issued a reset order for this specific attacker and will not re-issue one, and at each receiving node, which records that it has already executed the reset for this specific attacker and will not repeat it upon receiving a duplicate or late-arriving broadcast. The guard is enforced symmetrically at both ends because relying on either side alone is insufficient: a sender-only guard does not protect against duplicate messages already in flight through the gossip layer at the time the guard is set, and a receiver-only guard does not prevent the sender from continuing to needlessly re-broadcast on every subsequent round it still perceives \(n_{\text{adv}}\) as newly convicted. In early implementations that lacked this dual guard, reset broadcasts were observed recurring indefinitely, exceeding a thousand redundant broadcasts in a single run, which repeatedly re-triggered the weight reset at every receiving node and pinned the network's accuracy near that of a randomly initialized model for the remainder of the run. This underscores that the guard is not a minor implementation detail but a correctness requirement for the recovery protocol to function at all.

\subsection{Complete DecoyTrace Lifecycle}
\label{sec:methodology-lifecycle}

Algorithm~\ref{alg:DecoyTrace} formalizes the complete lifecycle of the system, integrating the five modules described in the preceding sections into a single flow executed by the DecoyNode in every audit round. The full pseudocode is presented, rather than a simplified summary, because the precise ordering and branching structure of Phase 4 is itself part of the paper's contribution, since the three control routes, malicious, carrier, and benign, are mutually exclusive by construction. The correctness of the hybrid gate depends on this exact evaluation order, as discussed below the algorithm.

\begin{algorithm}[h!]
\caption{DecoyTrace Lifecycle}
\label{alg:DecoyTrace}
\footnotesize
\begin{algorithmic}[1]
\Require Neighborhood $\mathcal{N}$, network participants $\mathcal{P}$, visited history $\mathcal{H}$, chaotic state $\sigma_t$, reputation module $\mathcal{R}$, thresholds $\tau_{\text{susp}}, \varepsilon, \tau_{\text{low-honest}}, \tau_{\text{honest}}, \rho, H$
\Ensure Control action $\mathcal{A}$
\State \textbf{// Phase 1--2: Bait generation and dissemination}
\State $\sigma_{t+1} \gets \texttt{LogisticMap}(\sigma_t)$; \ $\pi_t \gets \texttt{DeriveDecoyMap}(\sigma_{t+1})$ \Comment{Eq.~\ref{eq:logisticmap}--\ref{eq:honeymap}}
\State $\theta_{\text{bait}} \gets \texttt{FineTune}(\theta_{\text{global}}, \texttt{InjectTrigger}(\mathcal{D}_{\text{benign}}, \pi_t, p_{\text{inj}}))$ \Comment{Eq.~\ref{eq:trapdataset}}
\For{$n_j \in \mathcal{N}$}
    \State \texttt{Send}$(n_j, \Psi_{\text{tx}}(n_j))$ \Comment{Eq.~\ref{eq:dualmodel}}
\EndFor
\Statex
\State \textbf{// Phase 3: Semantic verification}
\For{$n_j \in \mathcal{N}$ such that $\theta_{\text{rx}}^{(n_j)}$ was received}
    \State $(CR, SR, HR) \gets \texttt{Infer}(\theta_{\text{rx}}^{(n_j)}, \mathcal{V}_{\text{trap}})$; \ \texttt{UpdateStreak}$(n_j, CR, SR, HR)$ \Comment{Eq.~\ref{eq:cr}--\ref{eq:local-streak}}
\EndFor
\Statex
\State \textbf{// Phase 4: Attribution and control branching}
\For{$n_j \in \mathcal{N}$}
    \If{$\texttt{local\_streak}(n_j) \geq \rho$}
        \If{$\texttt{is\_silent\_sink}(n_j, \mathcal{R})$} \Comment{Eq.~\ref{eq:silent-sink}}
            \State \Return \texttt{CONTAINMENT}$(n_j)$ \Comment{MALICIOUS: Phase 5}
        \ElsIf{$n_j \notin \mathcal{H}$}
            \State $\mathcal{H} \gets \mathcal{H} \cup \{n_j\}$; \ \texttt{Transfer}$(\sigma_t, \mathcal{H}, \theta_{\text{bait}}) \to n_j$
            \State \Return \texttt{PIVOT\_TO}$(n_j)$ \Comment{CARRIER: DFS pivot}
        \Else
            \State \Return \texttt{STAY} \Comment{Avoid revisiting cycles}
        \EndIf
    \ElsIf{$HR(n_j) \geq \tau_{\text{honest}}$}
        \State $\mathcal{T}(n_j) \gets \text{BENIGN}$ \Comment{BENIGN}
    \EndIf
\EndFor
\If{no carrier/attacker found in $\mathcal{N} \setminus \mathcal{H}$}
    \State $n^{*} \gets \texttt{NavigationTarget}(\mathcal{R}, \mathcal{H})$ \Comment{Eq.~\ref{eq:nav-target}}
    \If{$n^{*} \neq \emptyset$} \Return \texttt{PIVOT\_TO}$(\texttt{NextHopBFS}(n^{*}))$ \EndIf
\EndIf
\State \Return \texttt{STAY} \Comment{Continuous monitoring}
\Statex
\State \textbf{// Phase 5: Containment (invoked from line 12)}
\Procedure{Containment}{$n_{\text{adv}}$}
    \State $\texttt{targets} \gets \texttt{Neighbors}(n_{\text{adv}}, \text{global topology})$; \ \texttt{Connect}(\texttt{targets})
    \State \texttt{Broadcast}(\texttt{BLOCK\_NEIGHBOR}$(n_{\text{adv}})$, \texttt{targets})
    \If{$\texttt{global\_reset}$ enabled \textbf{and} $n_{\text{adv}} \notin \texttt{reset\_broadcast\_done}$}
        \State \texttt{Broadcast}(\texttt{MODEL\_RESET}$(\texttt{round}_{\text{current}}{+}1)$, $\mathcal{P}$) \Comment{Sender guard}
        \State $\texttt{reset\_broadcast\_done} \gets \texttt{reset\_broadcast\_done} \cup \{n_{\text{adv}}\}$
    \EndIf
\EndProcedure
\end{algorithmic}
\end{algorithm}

Three mutually exclusive control routes are distinguished in Phase 4, evaluated in a specific, non-arbitrary order, and correspond directly to the STAY, PIVOT, and CONTAIN control states shown in the legend of Figure~\ref{fig:architecture}. The first route, \textit{malicious}, applies when the sustained local signature matches the silent-sink profile, thereby immediately triggering containment. This check is evaluated first because it is the only route with an irreversible, network-wide side effect, and prioritizing it ensures the attacker is isolated as soon as both conditions of the hybrid gate are met, without waiting for the evaluation of the other neighbors in the same round. The second route, \textit{carrier}, applies when a local signature exists without attribution corroboration, in which case the auditor role recursively pivots toward the suspect to continue tracing upstream. The explicit history check, \(n_j \notin \mathcal{H}\), at this point is what prevents the DFS from re-entering a node it has already visited and looping indefinitely between two or more carriers that both continue to exhibit a residual signature after the DecoyNode has passed through them. The third route, \textit{benign}, applies when the honesty rate exceeds the threshold \(\tau_{\text{honest}} = 0.80\) after a minimum number of rounds of evidence, updating the neighbor's reputation to \texttt{BENIGN} so that it receives the clean model \(\theta_{\text{clean}}\) going forward, moving it out of the pool of nodes that must be re-probed with bait in subsequent rounds. When no direct neighbor satisfies the pivot or containment conditions, the system falls back to distance-guided navigation, in line 24, to steer the search toward the most plausible suspect according to the global reputation graph, guaranteeing convergence of the trace even in topologies where the attacker is not directly reachable from the DecoyNode's current position within a single hop.

The algorithm's per-round complexity has direct implications for the itinerant single-agent design choice in Section~\ref{sec:methodology-overview}. Phases 1 through 3 are \(O(|\mathcal{N}|)\) in the size of the DecoyNode's immediate neighborhood, requiring one bait dissemination and one inference pass per neighbor. Phase 4's attribution check is \(O(1)\) per neighbor, given the reputation module's precomputed accusation counts. The navigation fallback is a single breadth-first search over the known topology, \(O(|V| + |E|)\) in the worst case, invoked only in the rare rounds where no direct neighbor yields a pivot or conviction target. The dominant recurring cost is therefore local to the DecoyNode's current neighborhood rather than global to the network, which makes the itinerant, single-active-agent design tractable at the topology sizes evaluated in Section~\ref{sec:experimental-matrix}.

\section{Experimental Setup}
\label{sec:experimental-setup}

This section describes the experimental design used to evaluate the viability and resilience of \textit{DecoyTrace} across a diverse and controlled range of network conditions, and it provides the configuration details necessary to reproduce every result reported in Section~\ref{sec:results}. All experiments are executed on the NEBULA platform~\cite{nebula}, which orchestrates each federation as a set of isolated Docker containers, one per participant plus one controller and frontend process, communicating over a dedicated virtual network. Every participant runs an independent instance of the training and defense stack described in Section~\ref{sec:methodology}, and no participant shares memory or process-level state with any other beyond the exchanged model updates and control messages defined by the protocol itself. All experiments are trained on CPU only, without GPU acceleration; this reflects a deployment choice rather than a hardware constraint, since it keeps the evaluation environment uniform and inexpensive to reproduce, and it is consistent with the modest per-round compute footprint of the shallow architectures described in Section~\ref{sec:experimental-federation}. Every experiment runs on a single physical host equipped with a 36-thread Intel Xeon E5-2697 v4 CPU (2.30\,GHz) and 62\,GB of RAM, with ten participant containers of a given configuration co-located on this host and scheduled by the Docker daemon; no configuration in the matrix saturated the available memory, and the operating system scheduler shared CPU time across containers rather than statically partitioning it.

A full factorial matrix isolating three independent variables of interest, dataset heterogeneity, topological connectivity, and the presence and configuration of the defense mechanism, structures the evaluation. This factorial design is necessary rather than incidental to the paper's claims, because Section~\ref{sec:methodology-attribution} shows that local semantic metrics alone are structurally insufficient for attribution in networks with iterative aggregation, so the empirical validation must in turn confirm that the attribution mechanism holds not just under one favorable topology, but across the connectivity spectrum in which the underlying failure mode, semantic-fingerprint propagation to honest neighbors, actually arises.

\subsection{Experimental Matrix: Datasets, Topologies, Scenarios, and Adversary}
\label{sec:experimental-matrix}

Every combination of three independent axes is evaluated, yielding a total of \(5 \times 3 \times 4 = 60\) distinct experimental configurations.

\subsubsection{Datasets}
\label{sec:experimental-datasets}

Five image-classification benchmarks are evaluated, spanning a range of class cardinality, input modality, and task difficulty: MNIST~\cite{lecun1998mnist}, FashionMNIST~\cite{xiao2017fashionmnist}, and EMNIST~\cite{cohen2017emnist} (grayscale, \(|\mathcal{Y}| \in \{10, 10, 47\}\) respectively), and CIFAR-10~\cite{krizhevsky2009cifar} and CIFAR-100~\cite{krizhevsky2009cifar} (RGB, \(|\mathcal{Y}| \in \{10, 100\}\)). This range is deliberate along two axes. First, EMNIST and, more markedly, CIFAR-100 stress-test the decoy\_map permutation of Eq.~\ref{eq:honeymap} and the \(SR\) metric of Eq.~\ref{eq:sr}, since both are defined over the full label space \(\mathcal{Y}\); a mechanism validated only on 10-class datasets would leave open the question of whether attribution remains reliable as the space of plausible attack-target classes grows by an order of magnitude, because a larger \(|\mathcal{Y}|\) both dilutes the per-class probability mass available to \(SR\)'s \(\arg\max\) and increases the number of directions along which an attacker's poisoned prediction could plausibly drift. Second, task difficulty varies substantially across this set, since the severity of the attack's impact on honest-node utility is observed empirically (Section~\ref{sec:results}) to scale with intrinsic task difficulty, ranging from near-negligible degradation on MNIST to substantial degradation on CIFAR-10 and CIFAR-100. This allows \textit{DecoyTrace} to be evaluated both in the regime where the attack is easy to tolerate and in the regime where defense is most needed. For every dataset, samples are partitioned across participants in a non-independent and identically distributed (Non-IID~\cite{zhao2018noniid}) fashion using a Dirichlet distribution~\cite{hsu2019dirichlet} with concentration parameter \(\alpha = 0.5\), which induces a severe semantic skew in which each node holds examples predominantly from a minority of classes. This heterogeneity forces a natural divergence in local gradients \(\nabla w\) across honest nodes, and this divergence is precisely the regime in which passive defenses based on parameter-space distance, such as cosine similarity or Euclidean distance, lose discriminative power, so it is the regime in which the behavioral verification of Section~\ref{sec:methodology-detection} must be shown to hold.

\subsubsection{Topologies}
\label{sec:experimental-topologies}

Topological connectivity is varied instead of fixed to a single graph: Erd\H{o}s R\'enyi~\cite{erdos1959random} random graphs are generated at three edge probabilities, \(edge \in \{0.2, 0.5, 1.0\}\), over \(N = 10\) participants. This choice follows directly from the pivoting mechanism of Section~\ref{sec:methodology-pivot}. An edge probability of \(p = 0.2\) produces a sparse, high-diameter topology in which the attacker may be several hops away from the DecoyNode's initial position and may have only a single honest neighbor capable of accusing it, so it exercises both the distance-guided navigation of Eq.~\ref{eq:nav-target} and the \(\text{accused\_by} \geq 1\) design decision defended in Section~\ref{sec:methodology-attribution}. An edge probability of \(p = 1.0\) produces a fully connected topology in which every node is a direct neighbor of every other, and this instead stresses the semantic-fingerprint-propagation failure mode itself, since every honest node aggregates directly with the attacker every round. The intermediate value \(p = 0.5\) provides a density between these two extremes. Because a randomly sampled Erd\H{o}s R\'enyi graph is not guaranteed to be connected, especially at \(p = 0.2\), the graph is regenerated until network-wide connectivity is verified, so a disconnected topology is never silently evaluated as if it were a connectivity failure of the defense mechanism itself rather than of the underlying graph sample. A single random seed is not fixed for topology generation across repetitions of the same dataset and scenario configuration, so the specific graph instance realized at \(edge=0.2\) and \(edge=0.5\) varies across runs within the constraints above. This choice is deliberate, since it avoids overfitting the reported results to a single favorable or unfavorable graph sample, although it means that the topology itself is not independently reproducible bit-for-bit across re-runs, unlike data partitioning and model initialization, which are seeded and reproducible, as described below.

\subsubsection{Scenarios}
\label{sec:experimental-scenarios}

Four scenarios isolate the contribution of each system component for every combination of dataset and topology. In the Benign scenario, no adversarial participant is present, so this scenario establishes the utility ceiling of the federation, i.e., the accuracy and convergence behavior achievable in the complete absence of an attack, and serves as the reference against which the utility costs of both the attack and the defense are measured. In the Attack-only scenario, a single participant executes the semantic label-flipping attack described below with no defense mechanism active. This scenario establishes the utility floor, quantifying the damage an undefended network sustains, and serves as the baseline against which the recovery scenarios below are compared. In DecoyTrace, No Global Reset scenario, the full detection, attribution, and containment pipeline of Sections~\ref{sec:methodology-detection} through \ref{sec:methodology-pivot} is active, and upon conviction the network executes only the isolation phase through the \texttt{BLOCK\_NEIGHBOR} message, continuing training without a coordinated weight reset. In DecoyTrace, Global Reset scenario, the configuration is identical to the previous one, except that conviction additionally triggers the coordinated global reset through the \textit{SYNC\_SEED} mechanism described in Section~\ref{sec:methodology-pivot}.

Both recovery configurations are evaluated, rather than only the strategy that performs best in aggregate, because Section~\ref{sec:methodology-pivot} explicitly frames the choice between them as a trade-off rather than a dominance relation: Global Reset guarantees complete removal of residual poisoning at the cost of discarding accumulated training progress, whereas No Global Reset preserves progress at the cost of a decaying residual bias. Evaluating only one configuration would leave that trade-off as an unverified design claim rather than a measured result. Running Benign and Attack-only across the same three topology densities as the two DecoyTrace configurations, rather than fixing them to a single topology, is likewise deliberate, since it allows observed differences in the DecoyTrace scenarios to be attributed specifically to the defense mechanism, rather than to topology-dependent variation in the underlying federation dynamics that would also be present without any defense at all.

A fixed random seed, shared across all four scenarios within a given combination of dataset and topology, controls data partitioning and model initialization, so differences in outcome are attributable to the scenario condition itself, namely the presence and configuration of the attacker and of the defense, rather than to sampling variance in which classes each node happens to hold.

\subsubsection{Adversary}
\label{sec:experimental-adversary}

Consistent with the threat model in Section~\ref{sec:methodology-threat-model}, exactly one participant is designated as the attacker in every Attack-only, DecoyTrace No Global Reset, and DecoyTrace Global Reset run, and this participant is consistently identified across all configurations of a given topology as participant index 2 of the ten participants. The attacker executes a targeted semantic label-flipping attack, training on a locally relabeled copy of its own partition that redirects a chosen source class toward an attacker-selected target class, and submits the resulting update to its neighbors as any other participant would, with no parameter-space camouflage beyond what naturally results from FedAvg-style local training. The attacker's update norm or cosine similarity to the honest distribution is not additionally constrained in these experiments. As discussed in Section~\ref{sec:methodology-threat-model}, \textit{DecoyTrace}'s behavioral verification is designed to be agnostic to such camouflage, and evaluating the undefended, unconstrained attack first establishes the baseline severity of the threat before any defense-specific robustness claim is made.

\subsection{Federation Protocol and Training Configuration}
\label{sec:experimental-federation}

Every configuration runs for \(T = 100\) global rounds, a horizon chosen to be long enough to capture both the asymptotic convergence behavior of the federation and the full temporal signature required by the sustained-streak detection criterion of Eq.~\ref{eq:local-streak}, where \(\rho = 4\) rounds, including the subsequent recovery trajectory after conviction. FedAvg is used as the consensus function, preferred over aggregation rules with native Byzantine-fault tolerance, such as Krum~\cite{blanchard2017krum} or coordinate-wise trimmed mean~\cite{yin2018trimmedmean}, for a deliberate and analytical reason: because FedAvg is a purely linear averaging rule with no statistical outlier defense of its own, any resilience observed in the DecoyTrace scenarios can be attributed specifically to the semantic detection and attribution logic injected by the DecoyNode, rather than to a confound from the aggregation rule's own robustness properties. This choice isolates the mechanism's contribution and, incidentally, represents the more demanding evaluation setting, since a defense that is effective on top of an undefended aggregator constitutes a stronger result than one whose apparent effectiveness might otherwise be inseparable from the aggregator's own fault tolerance.

Each participant performs a single local training epoch per global round before submitting its update, using the Adam optimizer throughout. For MNIST, FashionMNIST, EMNIST, and CIFAR-10, all participants share a fixed learning rate of \(10^{-3}\) with default Adam momentum coefficients. For CIFAR-100, given its substantially larger label space and correspondingly harder optimization landscape, a separately tuned Adam configuration is used, with learning rate \(8.05 \times 10^{-5}\), \(\beta_1 = 0.8514\), and \(\beta_2 = 0.9997\), with AMSGrad enabled, held fixed and identical across all CIFAR-100 scenarios and topologies. This is the only per-dataset deviation from an otherwise uniform training configuration, and because it is applied uniformly across every CIFAR-100 scenario, no cross-scenario comparison within that dataset is confounded by it. The model architecture is matched to each dataset's input modality: a multilayer perceptron for the grayscale datasets, namely MNIST, FashionMNIST, and EMNIST, and a convolutional neural network for the RGB datasets, namely CIFAR-10 and CIFAR-100. Model capacity, meaning input layer width and output layer width, is adjusted only to each dataset's native input dimensionality and class cardinality \(|\mathcal{Y}|\), and it is never adjusted to the experimental scenario. All experiments share identical decoy hyperparameters to those defined in Section~\ref{sec:methodologybait}, namely \(r = 3.99\), a \(4 \times 4\) trigger patch, and \(p_{\text{inj}} = 0.5\), and to those defined in Section~\ref{sec:methodology-detection}, namely \(\tau_{\text{susp}} = 0.40\), \(\varepsilon = 0.02\), \(\tau_{\text{low-honest}} = 0.35\), and \(\rho = 4\), so no observed difference across datasets or topologies can be attributed to an undocumented per-run hyperparameter change.

\section{Results}
\label{sec:results}

This section presents the empirical evaluation of \textit{DecoyTrace} across the full experimental matrix described in Section~\ref{sec:experimental-matrix}. Each of the following subsections is dedicated to a single dataset and reports two complementary views of the same evidence. The per-participant F1-score~\cite{vanrijsbergen1979ir} trajectory over the full \(T = 100\)-round horizon, broken down by topology and scenario, is reported first for each dataset, since it exposes the round-by-round dynamics that a single aggregate number cannot, including convergence speed, per-node divergence under attack, and the shape of the post-conviction recovery trajectory. The steady-state quantitative summary is reported secondly, giving a single comparable number per scenario alongside its process-level resource footprint. This is computed as the mean and standard deviation over the final 30 rounds of each honest participant and aggregated across the ten (or fewer, once the attacker or the decoy itself is excluded) trajectories. Each result below reflects the four-scenario protocol of Section~\ref{sec:experimental-scenarios}: \textit{Baseline} establishes the utility ceiling in the complete absence of an attacker; \textit{Attack} establishes the utility floor under an undefended poisoning attempt; and the two DecoyTrace configurations, with and without coordinated global reset, quantify the recovery achieved once detection, attribution, and containment are active.

Five datasets are analyzed here: MNIST, FashionMNIST, EMNIST, CIFAR 10, and CIFAR 100. The first three share the multilayer perceptron configuration detailed in the referenced section, isolating the label space effect and task difficulty without architectural confounds, whereas CIFAR 10 and CIFAR 100 employ a convolutional architecture over RGB inputs. Ordering these sets from the simplest classifier, MNIST, passing to another identical in size yet visually harder, FashionMNIST, then to a larger label space in EMNIST, and finally extending to the rich visual domains of CIFAR 10 and CIFAR 100, traces a complete difficulty gradient to compare both attack and defense directly.

\subsection{MNIST}
\label{sec:results-mnist}

\begin{figure}[t]
    \centering
    \includegraphics[width=\linewidth]{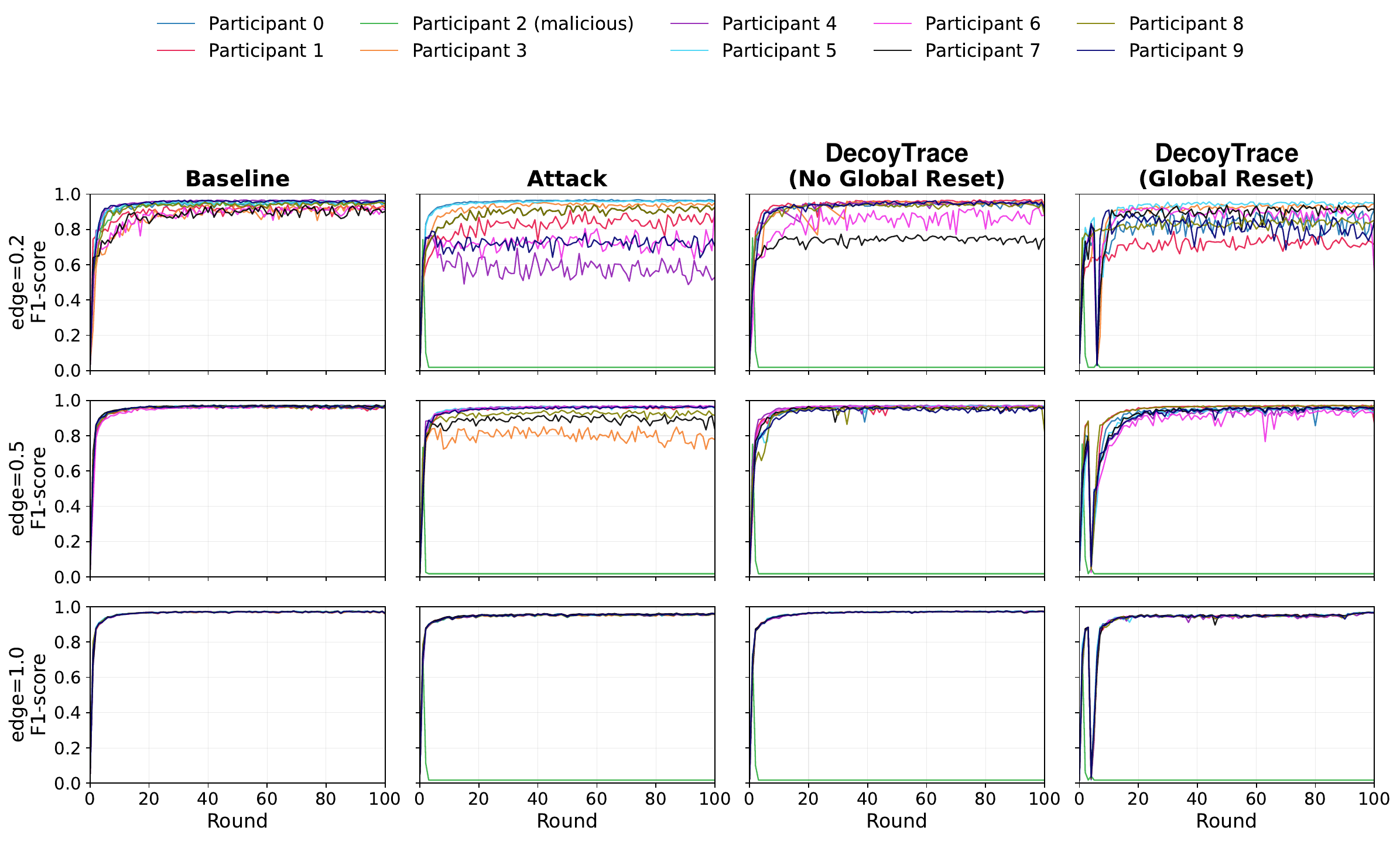}
    \caption{Per-participant F1-score over 100 rounds for MNIST, across the three evaluated topologies (rows, top to bottom: \(edge=0.2\), \(edge=0.5\), \(edge=1.0\)) and the four evaluated scenarios}
    \label{fig:f1_mnist}
\end{figure}

Figure~\ref{fig:f1_mnist} and Table~\ref{tab:summary_mnist} report the MNIST results. MNIST is the easiest classification task among the five datasets analyzed in this section, and the figure reflects this directly: every honest node in the Baseline column converges to a tight, near-ceiling F1-score band within the first 10 to 15 rounds, regardless of topology, consistent with the steady-state values in Table~\ref{tab:summary_mnist} ranging from \(0.936\) at \(edge=0.2\) to \(0.971\) at \(edge=1.0\). The Attack column shows that the cost of an undefended poisoning attempt scales inversely with topology density. At \(edge=0.2\), the sparse topology, honest-node F1-score drops to \(0.838\) with a wide standard deviation of \(0.134\), visible in the figure as several honest trajectories that separate from the tight Baseline band and oscillate before partially recovering as the poison dilutes through repeated aggregation. At \(edge=0.5\) and \(edge=1.0\), the degradation is markedly smaller, \(0.931\) and \(0.956\) respectively, because a denser topology dilutes the attacker's single poisoned contribution across a proportionally larger number of averaging partners at every round, so its per-round influence on any individual honest node's aggregate is weaker even though the attacker reaches more neighbors directly.

The two DecoyTrace columns show that both recovery strategies restore utility substantially above the undefended Attack scenario at every topology, and the No Global Reset configuration in particular nearly closes the gap to Baseline: \(0.913\) versus \(0.936\) at \(edge=0.2\), \(0.961\) versus \(0.964\) at \(edge=0.5\), and \(0.972\) versus \(0.971\) at \(edge=1.0\), the last a case in which the defended network's steady-state F1 marginally exceeds the undefended Baseline itself, within the margin of run-to-run variance given the unfixed topology sampling described in Section~\ref{sec:experimental-topologies}. The Global Reset configuration recovers less fully than No Global Reset at \(edge=0.2\) and \(edge=1.0\), \(0.857\) and \(0.954\) respectively, which it is the expected signature of the training-progress rollback discussed in Section~\ref{sec:methodology-pivot}: restoring all honest weights to \textit{SYNC\_SEED} discards accumulated convergence, so the post-reset trajectory visible in the figure has to re-climb from a lower point within the remaining rounds of the horizon, whereas No Global Reset only has to dilute a bounded residual bias out of weights that were never discarded.

\begin{table*}[!hb]
\caption{Quantitative summary for MNIST across all evaluated topologies (mean \(\pm\) standard deviation). For scenarios with an attacker, performance statistics are computed exclusively over honest nodes.}
\label{tab:summary_mnist}
\centering
\small
\setlength{\tabcolsep}{5pt}
\resizebox{\textwidth}{!}{
\begin{tabular}{clccccc}
\toprule
\textbf{Topology} & \textbf{Scenario} & \textbf{F1-score} & \textbf{CPU (\%)} & \textbf{RAM (MB)} & \textbf{$\Delta$TX (MB/round)} & \textbf{$\Delta$RX (MB/round)} \\
\midrule
\multirow{4}{*}{$edge=0.2$}
 & Baseline (10 honest) & $\mathbf{0.936 \pm 0.024}$ & $87.51 \pm 72.62$ & $\mathbf{1128.76 \pm 33.89}$ & $0.619 \pm 1.075$ & $0.619 \pm 0.852$ \\
 & Attack (9 honest) & $0.838 \pm 0.134$ & $87.36 \pm 69.00$ & $1140.17 \pm 31.00$ & $0.797 \pm 1.323$ & $0.800 \pm 0.947$ \\
 & DecoyTrace (No Global Reset) (8 honest) & $0.913 \pm 0.071$ & $59.81 \pm 73.70$ & $1142.07 \pm 48.01$ & $0.368 \pm 0.826$ & $0.367 \pm 0.601$ \\
 & DecoyTrace (Global Reset) (8 honest) & $0.857 \pm 0.078$ & $\mathbf{47.18 \pm 70.02}$ & $1203.01 \pm 42.84$ & $\mathbf{0.273 \pm 0.786}$ & $\mathbf{0.275 \pm 0.515}$ \\
\midrule
\multirow{4}{*}{$edge=0.5$}
 & Baseline (10 honest) & $\mathbf{0.964 \pm 0.006}$ & $89.25 \pm 64.08$ & $\mathbf{1137.99 \pm 36.23}$ & $1.383 \pm 2.131$ & $1.383 \pm 1.349$ \\
 & Attack (9 honest) & $0.931 \pm 0.058$ & $97.72 \pm 61.82$ & $1145.23 \pm 33.29$ & $1.328 \pm 1.847$ & $1.329 \pm 1.221$ \\
 & DecoyTrace (No Global Reset) (8 honest) & $0.961 \pm 0.013$ & $33.85 \pm 58.63$ & $1169.98 \pm 47.19$ & $0.358 \pm 1.088$ & $0.362 \pm 0.676$ \\
 & DecoyTrace (Global Reset) (8 honest) & $0.953 \pm 0.016$ & $\mathbf{28.21 \pm 53.44}$ & $1187.87 \pm 40.57$ & $\mathbf{0.299 \pm 1.022}$ & $\mathbf{0.303 \pm 0.583}$ \\
\midrule
\multirow{4}{*}{$edge=1.0$}
 & Baseline (10 honest) & $0.971 \pm 0.003$ & $104.25 \pm 49.69$ & $\mathbf{1152.46 \pm 41.23}$ & $2.508 \pm 3.619$ & $2.510 \pm 2.541$ \\
 & Attack (9 honest) & $0.956 \pm 0.003$ & $99.86 \pm 50.98$ & $1163.86 \pm 36.72$ & $2.596 \pm 3.624$ & $2.602 \pm 2.538$ \\
 & DecoyTrace (No Global Reset) (8 honest) & $\mathbf{0.972 \pm 0.002}$ & $98.51 \pm 56.16$ & $1207.58 \pm 51.77$ & $2.141 \pm 3.394$ & $2.161 \pm 1.925$ \\
 & DecoyTrace (Global Reset) (8 honest) & $0.954 \pm 0.008$ & $\mathbf{54.23 \pm 64.42}$ & $1248.53 \pm 51.26$ & $\mathbf{1.030 \pm 2.542}$ & $\mathbf{1.044 \pm 1.877}$ \\
\bottomrule
\end{tabular}}
\end{table*}

Table~\ref{tab:summary_mnist} additionally reports the resource cost of the defense. Both DecoyTrace configurations consistently reduce mean process CPU utilization relative to Baseline and Attack at \(edge=0.2\) and \(edge=0.5\), for instance \(28.21\%\) for Global Reset versus \(89.25\%\) for Baseline at \(edge=0.5\), because once a neighbor is convicted and blocked, the honest nodes adjacent to it stop expending compute on aggregating and validating updates that would otherwise still be arriving from that connection. This effect is markedly weaker at \(edge=1.0\), where CPU for DecoyTrace No Global Reset, \(98.51\%\), sits close to Baseline's \(104.25\%\), since every node is already a direct neighbor of every other in a fully connected topology, so blocking the single attacker removes only one connection out of nine rather than a proportionally larger share of a sparser neighborhood. Communication volume follows the same pattern for the same underlying reason: both \(\Delta\)TX and \(\Delta\)RX per round drop substantially in the DecoyTrace configurations relative to Baseline at every topology, most sharply at \(edge=1.0\), where DecoyTrace Global Reset's \(1.030\) MB/round in \(\Delta\)TX is under half of Baseline's \(2.508\), reflecting the bandwidth saved by no longer exchanging updates with a now-blocked neighbor in a topology where every edge otherwise carries traffic every round.

\subsection{FashionMNIST}
\label{sec:results-fashionmnist}

\begin{figure}[t]
    \centering
    \includegraphics[width=\linewidth]{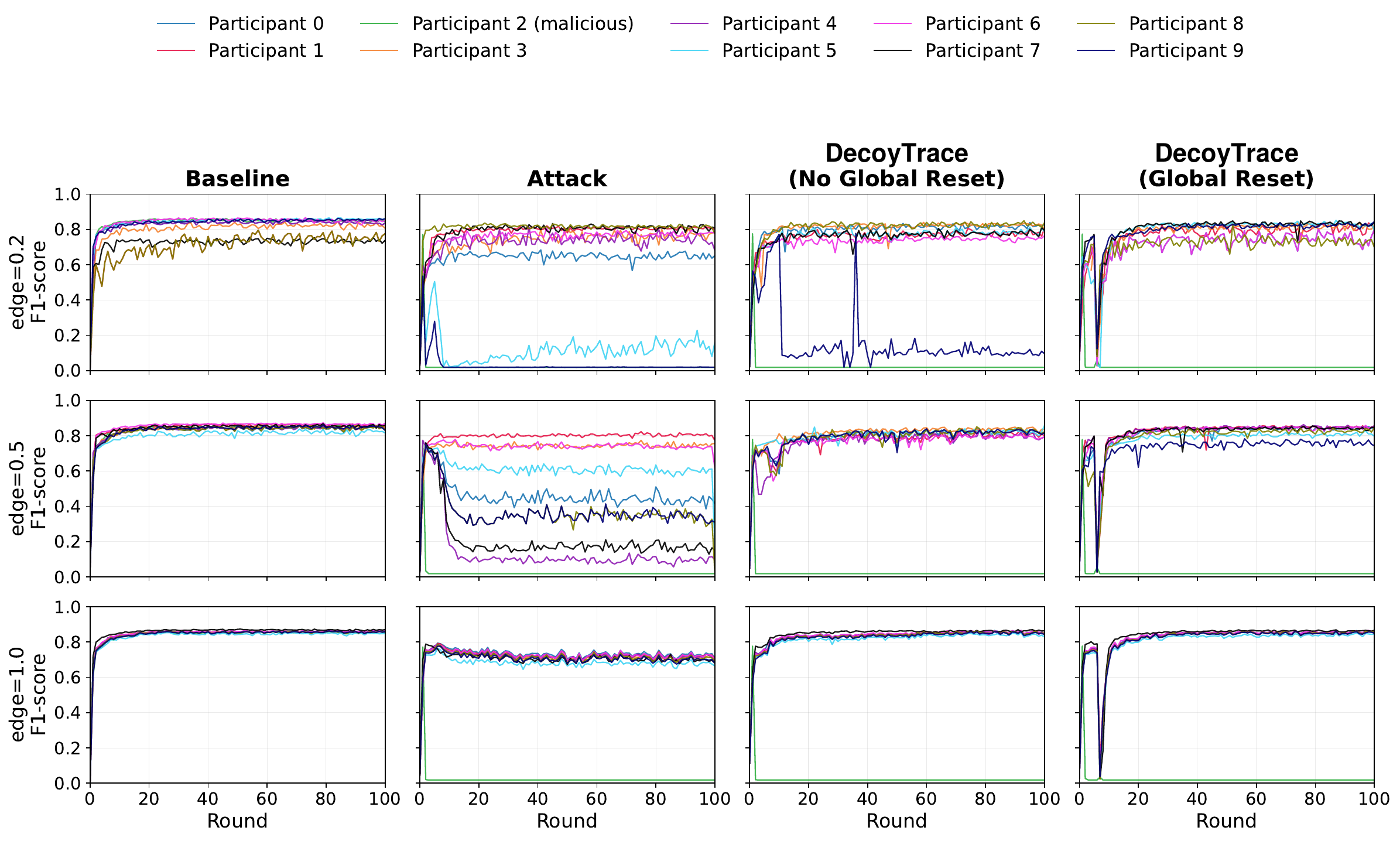}
    \caption{Per-participant F1-score over 100 rounds for FashionMNIST, across the three evaluated topologies (rows, top to bottom: \(edge=0.2\), \(edge=0.5\), \(edge=1.0\)) and the four evaluated scenarios}
    \label{fig:f1_fashionmnist}
\end{figure}

FashionMNIST shares MNIST's label cardinality and grayscale modality but poses a visually harder classification task, and Figure~\ref{fig:f1_fashionmnist} together with Table~\ref{tab:summary_fashionmnist} shows this immediately in a lower and noisier Baseline: \(0.815\) at \(edge=0.2\), \(0.848\) at \(edge=0.5\), and \(0.858\) at \(edge=1.0\), each with visibly wider inter-round oscillation in the figure than the corresponding MNIST panels. The Attack scenario is where the increased task difficulty becomes most consequential. At \(edge=0.5\), the undefended attack collapses honest-node F1-score to \(0.472\), a drop of over \(0.37\) from Baseline's \(0.848\) and the single largest attack-induced degradation observed among the three datasets in this section; the figure shows several honest trajectories at this topology failing to separate cleanly from the malicious one, unlike the corresponding MNIST panel where the attacker's own curve remains visually distinguishable throughout. \(edge=0.2\) shows a milder but still severe drop, to \(0.612\) with a very wide standard deviation of \(0.290\), reflecting the same sparse-topology partial-recovery pattern seen in MNIST. By contrast, \(edge=1.0\) degrades to \(0.709\), less severely than \(edge=0.5\) despite the denser connectivity, an inversion of the monotonic density-dilution trend observed in MNIST. This inversion is attributable to the specific random topology instance realized at that run rather than to a systematic property of edge density, consistent with the unseeded topology sampling described in Section~\ref{sec:experimental-topologies}.

The DecoyTrace configurations recover most, though not all, of this loss. At \(edge=0.5\), the topology with the most severe undefended degradation, No Global Reset restores F1-score to \(0.817\) and Global Reset to \(0.827\), both within \(0.03\) of Baseline's \(0.848\) despite starting from an Attack floor of \(0.472\), which is the clearest demonstration among the three datasets in this section of the defense's ability to recover from a severe attack rather than merely a mild one. At \(edge=1.0\), both configurations essentially match Baseline, \(0.854\) and \(0.853\) against \(0.858\). The \(edge=0.2\) topology is the one exception to this pattern. No Global Reset only reaches \(0.708\), below Global Reset's \(0.791\) and with a standard deviation of \(0.230\) that is nearly as wide as the undefended Attack scenario's own \(0.290\), visible in the figure as several honest trajectories at this topology that remain volatile well past the point of conviction. This is consistent with the sparse-topology attribution latency discussed in Section~\ref{sec:methodology-pivot}: at \(edge=0.2\) the attacker may have only a single honest neighbor capable of accusing it, so conviction and containment take longer to complete, and without a coordinated reset the residual bias absorbed by honest nodes before conviction has correspondingly less of the \(100\)-round horizon left to dilute away, whereas Global Reset removes that residual bias outright regardless of when conviction occurs.

\begin{table*}[!hb]
\caption{Quantitative summary for FashionMNIST across all evaluated topologies (mean \(\pm\) standard deviation). For scenarios with an attacker, performance statistics are computed exclusively over honest nodes.}
\label{tab:summary_fashionmnist}
\centering
\small
\setlength{\tabcolsep}{5pt}
\resizebox{\textwidth}{!}{
\begin{tabular}{clccccc}
\toprule
\textbf{Topology} & \textbf{Scenario} & \textbf{F1-score} & \textbf{CPU (\%)} & \textbf{RAM (MB)} & \textbf{$\Delta$TX (MB/round)} & \textbf{$\Delta$RX (MB/round)} \\
\midrule
\multirow{4}{*}{$edge=0.2$}
 & Baseline (10 honest) & $\mathbf{0.815 \pm 0.048}$ & $94.78 \pm 67.10$ & $1146.70 \pm 34.89$ & $0.760 \pm 1.262$ & $0.761 \pm 0.804$ \\
 & Attack (9 honest) & $0.612 \pm 0.290$ & $99.98 \pm 63.06$ & $\mathbf{1135.02 \pm 40.55}$ & $0.755 \pm 1.273$ & $0.753 \pm 0.971$ \\
 & DecoyTrace (No Global Reset) (8 honest) & $0.708 \pm 0.230$ & $\mathbf{24.23 \pm 51.48}$ & $1188.07 \pm 47.68$ & $\mathbf{0.178 \pm 0.650}$ & $\mathbf{0.179 \pm 0.407}$ \\
 & DecoyTrace (Global Reset) (8 honest) & $0.791 \pm 0.044$ & $37.85 \pm 63.14$ & $1170.83 \pm 44.07$ & $0.308 \pm 0.917$ & $0.311 \pm 0.631$ \\
\midrule
\multirow{4}{*}{$edge=0.5$}
 & Baseline (10 honest) & $\mathbf{0.848 \pm 0.013}$ & $93.86 \pm 59.80$ & $\mathbf{1147.38 \pm 38.92}$ & $1.672 \pm 2.422$ & $1.671 \pm 1.711$ \\
 & Attack (9 honest) & $0.472 \pm 0.247$ & $69.68 \pm 68.33$ & $1165.45 \pm 35.77$ & $1.064 \pm 2.021$ & $1.073 \pm 1.000$ \\
 & DecoyTrace (No Global Reset) (8 honest) & $0.817 \pm 0.018$ & $\mathbf{29.21 \pm 53.11}$ & $1181.26 \pm 42.48$ & $\mathbf{0.356 \pm 1.157}$ & $\mathbf{0.358 \pm 0.738}$ \\
 & DecoyTrace (Global Reset) (8 honest) & $0.827 \pm 0.030$ & $33.12 \pm 55.48$ & $1185.62 \pm 43.78$ & $0.426 \pm 1.325$ & $0.441 \pm 0.778$ \\
\midrule
\multirow{4}{*}{$edge=1.0$}
 & Baseline (10 honest) & $\mathbf{0.858 \pm 0.006}$ & $88.81 \pm 61.46$ & $\mathbf{1147.97 \pm 42.28}$ & $2.266 \pm 3.497$ & $2.286 \pm 2.181$ \\
 & Attack (9 honest) & $0.709 \pm 0.018$ & $\mathbf{87.42 \pm 58.08}$ & $1170.19 \pm 38.67$ & $2.124 \pm 3.417$ & $2.150 \pm 1.881$ \\
 & DecoyTrace (No Global Reset) (8 honest) & $0.854 \pm 0.008$ & $87.97 \pm 60.42$ & $1207.22 \pm 51.73$ & $\mathbf{1.865 \pm 3.248}$ & $\mathbf{1.897 \pm 1.557}$ \\
 & DecoyTrace (Global Reset) (8 honest) & $0.853 \pm 0.008$ & $91.22 \pm 58.30$ & $1213.99 \pm 56.35$ & $2.121 \pm 3.362$ & $2.163 \pm 1.423$ \\
\bottomrule
\end{tabular}}
\end{table*}

The resource-cost pattern observed for MNIST recurs here. CPU utilization for both DecoyTrace configurations drops sharply relative to Baseline and Attack at \(edge=0.2\) and \(edge=0.5\), for instance \(24.23\%\) for No Global Reset versus \(94.78\%\) for Baseline at \(edge=0.2\), and the gap narrows at \(edge=1.0\), where DecoyTrace CPU, \(87.97\%\) and \(91.22\%\), sits close to Baseline's \(88.81\%\), for the same fully-connected-topology reason discussed for MNIST. Communication volume mirrors this trend, most visibly at \(edge=0.2\), where DecoyTrace No Global Reset's \(\Delta\)TX of \(0.178\) MB/round is less than a quarter of Baseline's \(0.760\).

\subsection{EMNIST}
\label{sec:results-emnist}

\begin{figure}[t]
    \centering
    \includegraphics[width=\linewidth]{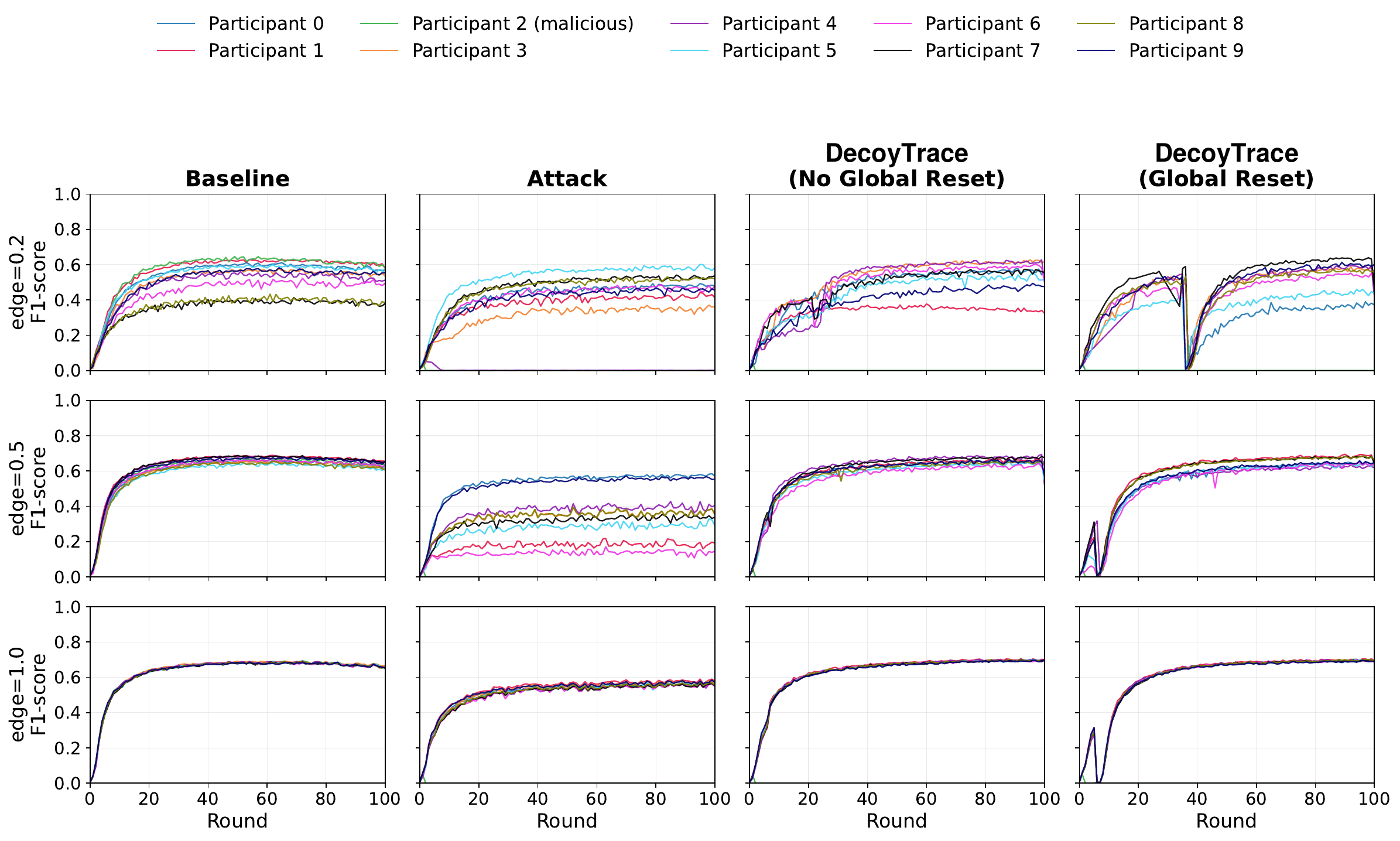}
    \caption{Per-participant F1-score over 100 rounds for EMNIST, across the three evaluated topologies (rows, top to bottom: \(edge=0.2\), \(edge=0.5\), \(edge=1.0\)) and the four evaluated scenarios.}
    \label{fig:f1_emnist}
\end{figure}

EMNIST expands the label space to \(|\mathcal{Y}| = 47\) classes, nearly five times MNIST and FashionMNIST, and Figure~\ref{fig:f1_emnist} together with Table~\ref{tab:summary_emnist} shows the effect of this expanded label space on every scenario simultaneously: Baseline F1-score is substantially lower in absolute terms than either of the other two datasets at every topology, \(0.530\) at \(edge=0.2\), \(0.648\) at \(edge=0.5\), and \(0.674\) at \(edge=1.0\), a direct consequence of the harder 47-way classification problem itself rather than of any attack or defense condition, since these are the scenarios in which no adversarial participant is present at all. The Attack scenario degrades this already-lower ceiling further, to \(0.421\), \(0.355\), and \(0.561\) respectively, and the relative drop at \(edge=0.5\), from \(0.648\) to \(0.355\), a reduction of nearly \(45\%\), is comparable in magnitude to the most severe degradation observed for FashionMNIST at the same topology, indicating that the attack's disruptive capacity does not diminish merely because the underlying task is already harder.

The DecoyTrace configurations show a distinctive pattern on EMNIST that differs qualitatively from both MNIST and FashionMNIST: the recovered F1-score not only approaches Baseline but exceeds it across every topology. No Global Reset reaches \(0.533\), \(0.652\), and \(0.694\) at \(edge=0.2\), \(edge=0.5\), and \(edge=1.0\) respectively, each above the corresponding Baseline value of \(0.530\), \(0.648\), and \(0.674\), and Global Reset follows the same pattern with \(0.529\), \(0.650\), and \(0.692\). This is visible in the figure as the DecoyTrace columns' honest trajectories settling at or slightly above the Baseline column's own band at every topology, most clearly at \(edge=1.0\). Given that Baseline involves no adversarial participant at all, this pattern cannot be read as the defense mechanism itself improving on an attack-free federation, since Baseline and the DecoyTrace scenarios differ in the composition of the aggregating subgraph (ten versus eight honest nodes) rather than only in the presence of the defense. A plausible, though not directly verified, hypothesis is that removing the single attacker's participant slot from the aggregation pool, once blocked via \texttt{BLOCK\_NEIGHBOR}, leaves the remaining eight honest nodes aggregating over a smaller but entirely clean set of updates in a Non-IID, 47-class setting where each participant's local class distribution is already sparse, so the ceiling achievable by this specific eight-node subgraph need not coincide with the ten-node Baseline ceiling and, in this label-space regime, could plausibly come out marginally higher. Confirming this hypothesis would require an additional ablation, namely re-running Baseline restricted to the same eight-node subset, which is left for future work. The magnitude of the gap in all cases, under \(0.02\) F1, is small relative to the topology-to-topology variation within a single scenario, so this pattern should be read as evidence that the defense fully neutralizes the attack's effect on convergence quality, rather than as a claim that defended federations systematically outperform undefended ones.

\begin{table*}[!hb]
\caption{Quantitative summary for EMNIST across all evaluated topologies (mean \(\pm\) standard deviation). For scenarios with an attacker, performance statistics are computed exclusively over honest nodes.}
\label{tab:summary_emnist}
\centering
\small
\setlength{\tabcolsep}{5pt}
\resizebox{\textwidth}{!}{
\begin{tabular}{clccccc}
\toprule
\textbf{Topology} & \textbf{Scenario} & \textbf{F1-score} & \textbf{CPU (\%)} & \textbf{RAM (MB)} & \textbf{$\Delta$TX (MB/round)} & \textbf{$\Delta$RX (MB/round)} \\
\midrule
\multirow{4}{*}{$edge=0.2$}
 & Baseline (10 honest) & $0.530 \pm 0.079$ & $66.49 \pm 68.75$ & $\mathbf{1210.61 \pm 48.50}$ & $0.203 \pm 0.664$ & $0.206 \pm 0.436$ \\
 & Attack (9 honest) & $0.421 \pm 0.161$ & $107.19 \pm 51.82$ & $1219.87 \pm 35.21$ & $0.404 \pm 1.010$ & $0.404 \pm 0.806$ \\
 & DecoyTrace (No Global Reset) (8 honest) & $\mathbf{0.533 \pm 0.085}$ & $\mathbf{55.35 \pm 67.05}$ & $1257.77 \pm 46.06$ & $\mathbf{0.179 \pm 0.664}$ & $\mathbf{0.184 \pm 0.414}$ \\
 & DecoyTrace (Global Reset) (8 honest) & $0.529 \pm 0.084$ & $58.40 \pm 67.33$ & $1235.61 \pm 39.70$ & $0.188 \pm 0.613$ & $0.188 \pm 0.417$ \\
\midrule
\multirow{4}{*}{$edge=0.5$}
 & Baseline (10 honest) & $0.648 \pm 0.018$ & $99.41 \pm 55.26$ & $\mathbf{1226.98 \pm 50.89}$ & $0.743 \pm 1.818$ & $0.756 \pm 1.204$ \\
 & Attack (9 honest) & $0.355 \pm 0.138$ & $98.61 \pm 56.42$ & $1234.20 \pm 39.28$ & $0.618 \pm 1.568$ & $0.629 \pm 1.116$ \\
 & DecoyTrace (No Global Reset) (8 honest) & $\mathbf{0.652 \pm 0.022}$ & $\mathbf{49.56 \pm 64.00}$ & $1256.77 \pm 40.23$ & $\mathbf{0.278 \pm 1.052}$ & $\mathbf{0.288 \pm 0.741}$ \\
 & DecoyTrace (Global Reset) (8 honest) & $0.650 \pm 0.023$ & $61.59 \pm 65.40$ & $1244.75 \pm 46.54$ & $0.426 \pm 1.415$ & $0.438 \pm 0.784$ \\
\midrule
\multirow{4}{*}{$edge=1.0$}
 & Baseline (10 honest) & $0.674 \pm 0.009$ & $90.33 \pm 59.66$ & $\mathbf{1237.10 \pm 57.71}$ & $0.927 \pm 2.495$ & $\mathbf{0.943 \pm 1.456}$ \\
 & Attack (9 honest) & $0.561 \pm 0.012$ & $108.76 \pm 45.13$ & $1239.06 \pm 47.31$ & $1.236 \pm 2.832$ & $1.249 \pm 1.939$ \\
 & DecoyTrace (No Global Reset) (8 honest) & $\mathbf{0.694 \pm 0.004}$ & $91.78 \pm 58.31$ & $1262.27 \pm 51.43$ & $\mathbf{0.921 \pm 2.474}$ & $0.952 \pm 1.393$ \\
 & DecoyTrace (Global Reset) (8 honest) & $0.692 \pm 0.005$ & $\mathbf{87.86 \pm 58.26}$ & $1275.02 \pm 51.87$ & $0.977 \pm 2.535$ & $1.010 \pm 1.347$ \\
\bottomrule
\end{tabular}}
\end{table*}

The resource-cost pattern differs from MNIST and FashionMNIST in one respect worth noting explicitly. At \(edge=1.0\), DecoyTrace No Global Reset's CPU utilization, \(91.78\%\), is close to Baseline's \(90.33\%\) and Attack's \(108.76\%\), consistent with the fully-connected-topology effect already discussed for the other two datasets, where blocking one attacker out of nine neighbors saves proportionally little compute. At \(edge=0.2\) and \(edge=0.5\), however, the CPU reduction is present but comparatively smaller in absolute terms than the corresponding MNIST and FashionMNIST reductions, for instance \(55.35\%\) for DecoyTrace No Global Reset against \(66.49\%\) for Baseline at \(edge=0.2\), a narrower gap than the roughly \(30\)-point gaps observed for the other two datasets at the same topology. This is consistent with the larger, 47-class MLP output layer required for EMNIST, described in Section~\ref{sec:experimental-federation}, increasing the fixed per-round inference and fine-tuning cost that every node, blocked-neighbor savings notwithstanding, still incurs regardless of scenario.

\subsection{CIFAR-10}
\label{sec:results-cifar10}

\begin{figure}[t]
    \centering
    \includegraphics[width=\linewidth]{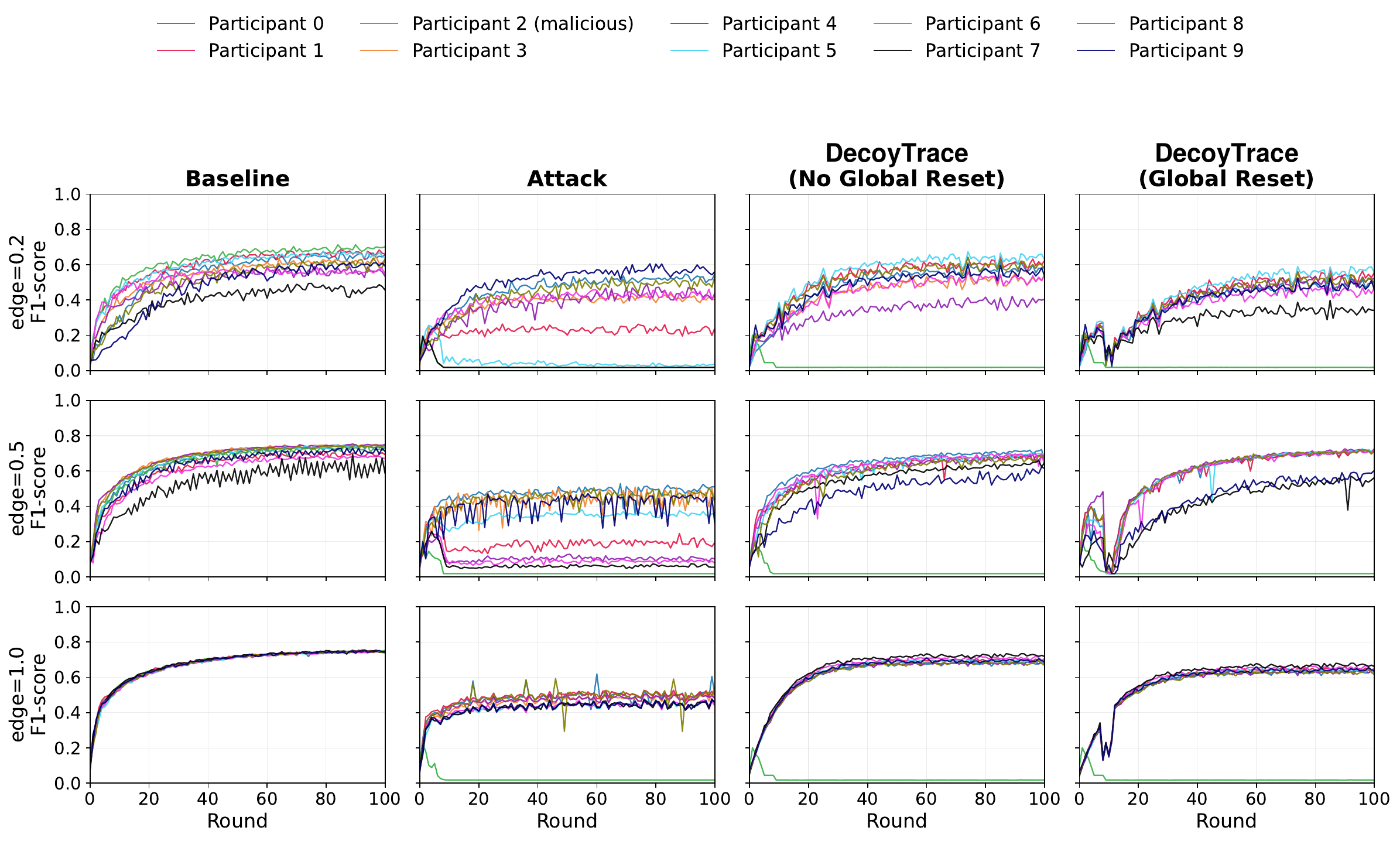}
    \caption{Per-participant F1-score over 100 rounds for CIFAR-10, across the three evaluated topologies (rows, top to bottom: \(edge=0.2\), \(edge=0.5\), \(edge=1.0\)) and the four evaluated scenarios.}
    \label{fig:f1_cifar10}
\end{figure}

CIFAR-10 shares MNIST's ten-class label space but replaces the grayscale multilayer perceptron with the convolutional architecture described in Section~\ref{sec:experimental-federation}, evaluated over RGB inputs of substantially higher visual complexity. Figure~\ref{fig:f1_cifar10} together with Table~\ref{tab:summary_cifar10} shows a Baseline ceiling clearly below MNIST's at every topology despite the identical class cardinality, \(0.605\) at \(edge=0.2\), \(0.712\) at \(edge=0.5\), and \(0.743\) at \(edge=1.0\), directly attributable to the harder underlying visual task rather than to the label space, which is the same ten classes evaluated for MNIST at a substantially higher ceiling. The Attack scenario severely degrades this ceiling across all topologies, dropping the honest-node F1-score to \(0.349\) at \(edge=0.2\), \(0.292\) at \(edge=0.5\), and \(0.472\) at \(edge=1.0\). The relative degradation at \(edge=0.5\), a drop of \(0.420\) from Baseline's \(0.712\), is the most severe undefended collapse observed across all four datasets analyzed in this section, exceeding even the most severe FashionMNIST degradation at the same topology, and the figure shows this visually as honest trajectories that fail to separate from the malicious participant's curve for a substantial portion of the round horizon before beginning to recover.

The two DecoyTrace configurations recover a substantial share of this loss, though the recovery is visibly less complete than for the three grayscale datasets. At \(edge=0.5\), the topology with the most severe undefended collapse, No Global Reset restores F1-score to \(0.659\) and Global Reset to \(0.665\), both closing most of the gap between the Attack floor of \(0.292\) and Baseline's \(0.712\), while remaining roughly \(0.05\) below Baseline itself rather than essentially matching it as observed for FashionMNIST at the same topology. At \(edge=1.0\), the pattern is similar: No Global Reset reaches \(0.694\) and Global Reset \(0.642\), both recovering the large majority of the gap from Attack's \(0.472\) to Baseline's \(0.743\) without fully closing it. At \(edge=0.2\), No Global Reset reaches \(0.547\) and Global Reset \(0.481\), again a substantial recovery from Attack's \(0.349\) that nonetheless leaves a wider residual gap to Baseline's \(0.605\) than in the corresponding MNIST or FashionMNIST panels. Across all three topologies, No Global Reset consistently outperforms Global Reset by a modest margin, the reverse of the ordering observed at several MNIST topologies, consistent with CIFAR-10's higher per-round training cost, described below, making the training-progress rollback inherent to \textit{SYNC\_SEED} comparatively more costly to recover from within the fixed \(100\)-round horizon.

\begin{table*}[!hb]
\caption{Quantitative summary for CIFAR-10 across all evaluated topologies (mean \(\pm\) standard deviation). For scenarios with an attacker, performance statistics are computed exclusively over honest nodes.}
\label{tab:summary_cifar10}
\centering
\small
\setlength{\tabcolsep}{5pt}
\resizebox{\textwidth}{!}{
\begin{tabular}{clccccc}
\toprule
\textbf{Topology} & \textbf{Scenario} & \textbf{F1-score} & \textbf{CPU (\%)} & \textbf{RAM (MB)} & \textbf{$\Delta$TX (MB/round)} & \textbf{$\Delta$RX (MB/round)} \\
\midrule
\multirow{4}{*}{$edge=0.2$}
 & Baseline (10 honest) & $\mathbf{0.605 \pm 0.064}$ & $201.04 \pm 111.19$ & $1379.80 \pm 36.60$ & $0.646 \pm 1.560$ & $0.647 \pm 1.267$ \\
 & Attack (9 honest) & $0.349 \pm 0.196$ & $170.55 \pm 127.12$ & $\mathbf{1388.67 \pm 30.66}$ & $0.694 \pm 1.986$ & $0.702 \pm 1.394$ \\
 & DecoyTrace (No Global Reset) (8 honest) & $0.547 \pm 0.075$ & $\mathbf{107.22 \pm 145.82}$ & $1397.97 \pm 41.38$ & $\mathbf{0.323 \pm 1.133}$ & $\mathbf{0.330 \pm 0.747}$ \\
 & DecoyTrace (Global Reset) (8 honest) & $0.481 \pm 0.066$ & $115.17 \pm 157.30$ & $1422.36 \pm 65.56$ & $0.348 \pm 1.460$ & $0.359 \pm 1.041$ \\
\midrule
\multirow{4}{*}{$edge=0.5$}
 & Baseline (10 honest) & $\mathbf{0.712 \pm 0.041}$ & $144.26 \pm 124.78$ & $1404.57 \pm 35.92$ & $1.285 \pm 3.449$ & $1.311 \pm 2.162$ \\
 & Attack (9 honest) & $0.292 \pm 0.170$ & $156.42 \pm 124.57$ & $\mathbf{1398.41 \pm 31.24}$ & $1.198 \pm 3.099$ & $1.209 \pm 2.150$ \\
 & DecoyTrace (No Global Reset) (8 honest) & $0.659 \pm 0.043$ & $117.44 \pm 151.15$ & $1434.93 \pm 42.23$ & $0.835 \pm 2.695$ & $0.852 \pm 1.732$ \\
 & DecoyTrace (Global Reset) (8 honest) & $0.665 \pm 0.071$ & $\mathbf{99.21 \pm 142.14}$ & $1427.77 \pm 47.12$ & $\mathbf{0.774 \pm 2.813}$ & $\mathbf{0.803 \pm 1.833}$ \\
\midrule
\multirow{4}{*}{$edge=1.0$}
 & Baseline (10 honest) & $\mathbf{0.743 \pm 0.005}$ & $169.85 \pm 86.18$ & $\mathbf{1442.02 \pm 42.58}$ & $2.921 \pm 6.632$ & $2.932 \pm 5.237$ \\
 & Attack (9 honest) & $0.472 \pm 0.032$ & $167.74 \pm 88.77$ & $1451.38 \pm 46.35$ & $2.781 \pm 6.499$ & $2.789 \pm 4.683$ \\
 & DecoyTrace (No Global Reset) (8 honest) & $0.694 \pm 0.016$ & $\mathbf{130.52 \pm 120.35}$ & $1458.48 \pm 66.09$ & $\mathbf{1.744 \pm 5.230}$ & $\mathbf{1.819 \pm 3.159}$ \\
 & DecoyTrace (Global Reset) (8 honest) & $0.642 \pm 0.014$ & $140.75 \pm 125.35$ & $1472.04 \pm 62.46$ & $2.082 \pm 5.679$ & $2.140 \pm 3.184$ \\
\bottomrule
\end{tabular}}
\end{table*}

Table~\ref{tab:summary_cifar10} shows a process CPU footprint markedly higher than the three grayscale datasets across every scenario, exceeding \(140\%\) for several Baseline and Attack rows, a direct consequence of the convolutional architecture's larger per-round compute cost relative to the multilayer perceptron used for MNIST, FashionMNIST, and EMNIST. Both DecoyTrace configurations still reduce CPU utilization relative to Baseline and Attack at every topology, for instance \(99.21\%\) for Global Reset against \(144.26\%\) for Baseline at \(edge=0.5\), following the same blocked-neighbor mechanism discussed for the other datasets, although the absolute reduction is smaller in relative terms than for MNIST given the higher fixed cost every node incurs regardless of scenario. Communication volume follows the expected pattern as well, dropping in both DecoyTrace configurations relative to Baseline at every topology, most sharply at \(edge=0.2\), where DecoyTrace No Global Reset's \(\Delta\)TX of \(0.323\) MB/round is under half of Baseline's \(0.646\). RAM usage is the one resource metric that increases rather than decreases under DecoyTrace at every topology, for instance \(1472.04\) MB for Global Reset against \(1442.02\) MB for Baseline at \(edge=1.0\), consistent with the additional memory footprint of maintaining the dual-model structure of Section~\ref{sec:methodologydual} on top of the already larger convolutional model, a cost that is present but comparatively harder to observe against the smaller memory footprint of the multilayer-perceptron datasets.

\subsection{CIFAR-100}
\label{sec:results-cifar100}

\begin{figure}[t]
    \centering
    \includegraphics[width=\linewidth]{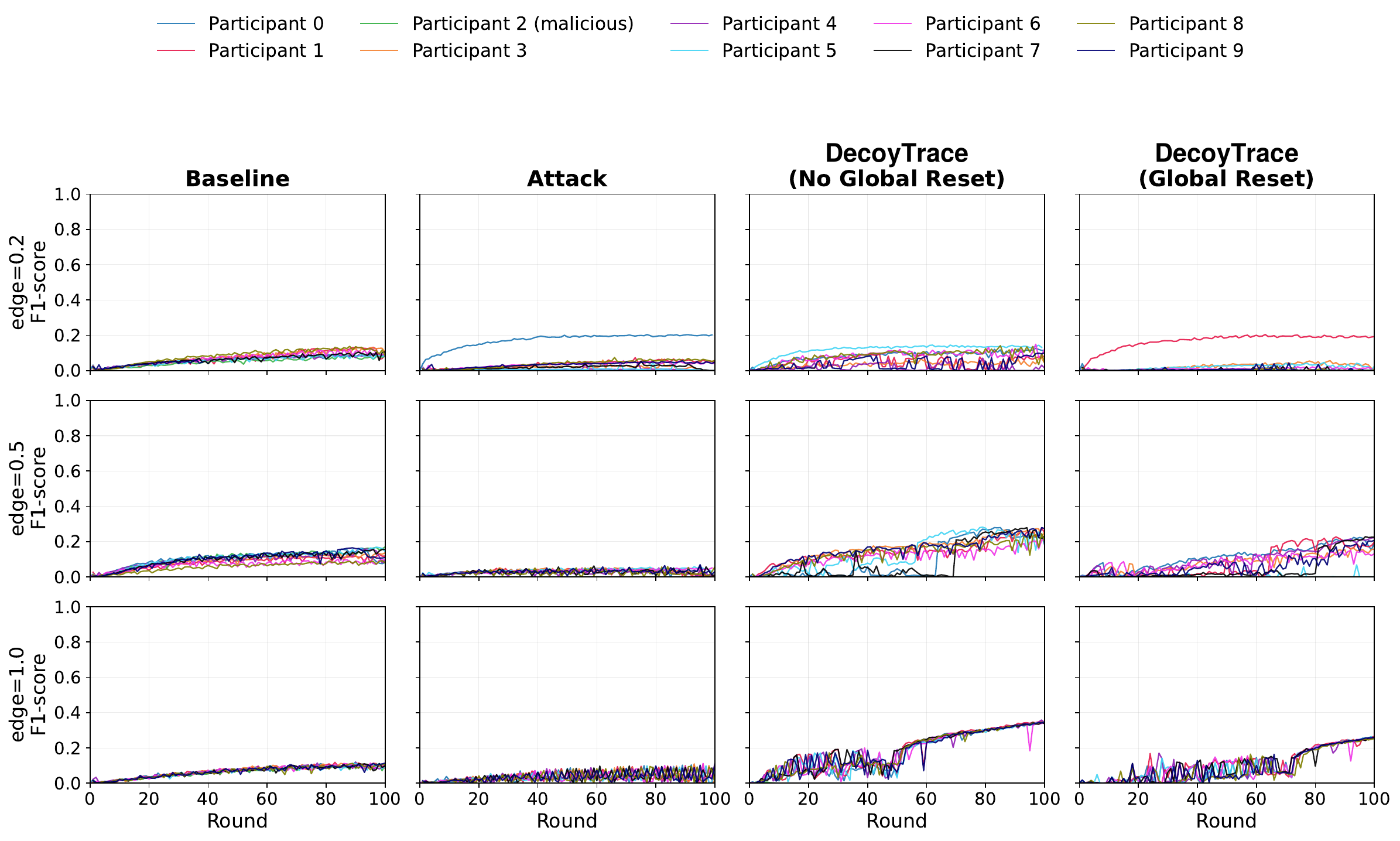}
    \caption{Per-participant F1-score over 100 rounds for CIFAR-100, across the three evaluated topologies (rows, top to bottom: \(edge=0.2\), \(edge=0.5\), \(edge=1.0\)) and the four evaluated scenarios.}
    \label{fig:f1_cifar100}
\end{figure}

CIFAR-100 expands the label space to \(|\mathcal{Y}| = 100\) classes, an order of magnitude beyond CIFAR-10 despite sharing the same convolutional architecture and RGB input modality, and Figure~\ref{fig:f1_cifar100} together with Table~\ref{tab:summary_cifar100} shows the effect of this expansion immediately: absolute F1-score collapses into a far lower regime than any of the four previously analyzed datasets, with Baseline itself only reaching \(0.090\) at \(edge=0.2\), \(0.120\) at \(edge=0.5\), and \(0.095\) at \(edge=1.0\). This low ceiling is a direct consequence of the 100-way classification problem's intrinsic difficulty under the Non-IID partitioning of Section~\ref{sec:experimental-datasets} rather than of any attack or defense condition, since these are the scenarios in which no adversarial participant is present at all, and the figure reflects it as visibly noisier, lower-amplitude honest trajectories across every panel compared to the tighter convergence bands seen for MNIST or CIFAR-10. The Attack scenario compresses this already-low ceiling further, to \(0.051\), \(0.030\), and \(0.047\) respectively, a relative reduction of roughly \(40\)-\(75\%\) depending on topology, showing that the attack remains disruptive in absolute terms even when the undefended ceiling it degrades is already close to the floor imposed by the task's own difficulty.

The two DecoyTrace configurations exhibit the most pronounced recovery-beyond-Baseline pattern observed across all five datasets analyzed in this section. No Global Reset reaches \(0.074\), \(0.218\), and \(0.313\) at \(edge=0.2\), \(edge=0.5\), and \(edge=1.0\) respectively, exceeding the corresponding Baseline value at \(edge=0.5\) and \(edge=1.0\) by a wide margin, more than doubling it at \(edge=0.5\) and more than tripling it at \(edge=1.0\); Global Reset follows a similar pattern at \(edge=0.5\) and \(edge=1.0\), reaching \(0.136\) and \(0.206\), though it remains below Baseline at \(edge=0.2\), at \(0.036\) against \(0.090\). This pattern is visible in the figure, with several DecoyTrace-column honest trajectories settling into a higher, less noisy F1-score band than their Baseline counterparts at the same topology, most clearly at \(edge=1.0\). As discussed for EMNIST in Section~\ref{sec:results-emnist}, this pattern should likewise not be read as the defense mechanism itself improving on an attack-free federation, since Baseline and the DecoyTrace scenarios again differ in the composition of the aggregating subgraph. The same eight-versus-ten-node hypothesis raised there remains the most plausible explanation, and it would be expected to apply with greater force here, since CIFAR-100's already-sparse per-class Non-IID partitioning at 100 classes could make the ceiling achievable by any given honest subgraph considerably more sensitive to which specific nodes compose it than in the ten-class datasets; under this hypothesis, removing the attacker's participant slot from the aggregation pool could shift the achievable ceiling by a much larger margin than the under-\(0.02\)-F1 shift observed for EMNIST. As with EMNIST, this explanation has not been isolated through a dedicated ablation and is reported here as a hypothesis rather than a demonstrated result. The exception at \(edge=0.2\) for Global Reset, where the training-progress rollback of \textit{SYNC\_SEED} is not fully recovered within the \(100\)-round horizon, is consistent with the same rollback cost discussed for CIFAR-10, compounded here by the intrinsically slower convergence of a 100-class problem.

\begin{table*}[!hb]
\caption{Quantitative summary for CIFAR-100 across all evaluated topologies (mean \(\pm\) standard deviation). For scenarios with an attacker, performance statistics are computed exclusively over honest nodes.}
\label{tab:summary_cifar100}
\centering
\small
\setlength{\tabcolsep}{5pt}
\resizebox{\textwidth}{!}{
\begin{tabular}{clccccc}
\toprule
\textbf{Topology} & \textbf{Scenario} & \textbf{F1-score} & \textbf{CPU (\%)} & \textbf{RAM (MB)} & \textbf{$\Delta$TX (MB/round)} & \textbf{$\Delta$RX (MB/round)} \\
\midrule
\multirow{4}{*}{$edge=0.2$}
 & Baseline (10 honest) & $\mathbf{0.090 \pm 0.025}$ & $363.94 \pm 185.64$ & $2486.50 \pm 590.10$ & $0.982 \pm 4.324$ & $0.966 \pm 3.087$ \\
 & Attack (9 honest) & $0.051 \pm 0.056$ & $344.42 \pm 153.35$ & $2132.93 \pm 561.91$ & $0.728 \pm 3.527$ & $0.726 \pm 2.571$ \\
 & DecoyTrace (No Global Reset) (8 honest) & $0.074 \pm 0.045$ & $381.00 \pm 197.44$ & $2696.47 \pm 541.82$ & $0.657 \pm 2.214$ & $0.695 \pm 2.070$ \\
 & DecoyTrace (Global Reset) (8 honest) & $0.036 \pm 0.061$ & $\mathbf{337.88 \pm 197.37}$ & $\mathbf{2102.09 \pm 538.20}$ & $\mathbf{0.550 \pm 1.563}$ & $\mathbf{0.566 \pm 1.842}$ \\
\midrule
\multirow{4}{*}{$edge=0.5$}
 & Baseline (10 honest) & $0.120 \pm 0.026$ & $344.42 \pm 178.06$ & $2374.68 \pm 704.47$ & $\mathbf{1.524 \pm 4.752}$ & $\mathbf{1.536 \pm 3.827}$ \\
 & Attack (9 honest) & $0.030 \pm 0.015$ & $\mathbf{336.59 \pm 189.68}$ & $2468.09 \pm 685.42$ & $1.929 \pm 3.258$ & $1.998 \pm 3.863$ \\
 & DecoyTrace (No Global Reset) (8 honest) & $\mathbf{0.218 \pm 0.041}$ & $373.73 \pm 305.65$ & $2349.71 \pm 554.94$ & $1.876 \pm 5.318$ & $2.103 \pm 3.864$ \\
 & DecoyTrace (Global Reset) (8 honest) & $0.136 \pm 0.069$ & $339.99 \pm 297.18$ & $\mathbf{2291.63 \pm 642.51}$ & $2.025 \pm 6.674$ & $2.443 \pm 4.673$ \\
\midrule
\multirow{4}{*}{$edge=1.0$}
 & Baseline (10 honest) & $0.095 \pm 0.010$ & $332.69 \pm 208.84$ & $3272.49 \pm 691.24$ & $14.235 \pm 9.636$ & $14.365 \pm 8.521$ \\
 & Attack (9 honest) & $0.047 \pm 0.033$ & $\mathbf{331.07 \pm 212.69}$ & $3178.18 \pm 665.34$ & $14.137 \pm 9.599$ & $14.180 \pm 8.479$ \\
 & DecoyTrace (No Global Reset) (8 honest) & $\mathbf{0.313 \pm 0.023}$ & $391.65 \pm 216.64$ & $\mathbf{3111.93 \pm 530.16}$ & $3.236 \pm 8.504$ & $3.722 \pm 7.103$ \\
 & DecoyTrace (Global Reset) (8 honest) & $0.206 \pm 0.052$ & $363.04 \pm 193.22$ & $3120.93 \pm 513.93$ & $\mathbf{3.096 \pm 8.250}$ & $\mathbf{3.551 \pm 7.073}$ \\
\bottomrule
\end{tabular}}
\end{table*}

Table~\ref{tab:summary_cifar100} shows the highest absolute resource footprint observed across all five datasets, with process CPU exceeding \(330\%\) in every single row and RAM reaching as high as \(3272\) MB for Baseline at \(edge=1.0\), reflecting the combined cost of the 100-class convolutional model's larger output layer and the separately tuned optimizer configuration described in Section~\ref{sec:experimental-federation}. Unlike the four smaller datasets, DecoyTrace's CPU reduction relative to Baseline and Attack is inconsistent here, and in several rows, such as DecoyTrace No Global Reset's \(391.65\%\) against Baseline's \(332.69\%\) at \(edge=1.0\), CPU is actually higher under the defense, which is attributable to the additional bait fine-tuning and semantic inference overhead of Section~\ref{sec:methodologybait} and Section~\ref{sec:methodology-detection} becoming proportionally more expensive to run atop an already computationally heavy 100-class model, an overhead that was comparatively negligible against the fixed cost of the smaller architectures used for the other datasets. Communication volume at \(edge=1.0\) is an order of magnitude higher than at the other two topologies for every scenario, exceeding \(14\) MB/round for both Baseline and Attack, a direct consequence of the fully connected topology multiplying the already large 100-class model's parameter count across nine simultaneous neighbor exchanges every round; DecoyTrace reduces this substantially at \(edge=1.0\), down to \(3.10\)-\(3.24\) MB/round for \(\Delta\)TX, the largest absolute bandwidth saving observed for any dataset in this section, since blocking the single attacker removes one of only ten total connections in a topology where every remaining edge still carries the full-sized 100-class model at every round.

\section{Discussion}
\label{sec:discussion}

The results in Section~\ref{sec:results} are analyzed here from a broader perspective. The goal is to move beyond the individual figures for each dataset to assess what they reveal collectively about \textit{DecoyTrace}. Furthermore, this approach is compared with the state-of-the-art reviewed in Section~\ref{sec:sota}, and its current limitations are discussed.

Across the five datasets, the three topologies, and both recovery configurations, \textit{DecoyTrace} consistently restores a substantial portion of the utility lost to an unprotected attack. In most configurations, it closes this gap almost entirely. On MNIST, the configuration without global reset achieves an F1-score just \(0.023\) below the baseline with \(edge=0.2\) and marginally exceeds it with \(edge=1.0\). This represents a recovery from a lower attack bound of just \(0.838\). FashionMNIST exhibits the same pattern under a substantially more severe attack. With \(edge=0.5\), the attack without defense causes the F1-score of honest nodes to plummet from a baseline of \(0.848\) to \(0.472\) (a drop of more than \(0.37\)). However, both DecoyTrace configurations recover to within \(0.03\) units of the baseline. EMNIST and CIFAR-100 go even further. The F1-score recovered by DecoyTrace exceeds the baseline in several topologies. This effect is particularly notable on CIFAR-100 with \(edge=1.0\), where the configuration without global reset reaches \(0.313\), compared to a baseline of only \(0.095\), thereby tripling the baseline. This latter pattern, detailed in Sections~\ref{sec:results-emnist} and \ref{sec:results-cifar100}, should not be interpreted as the defense mechanism itself improving upon an attack-free federation, since Baseline and the DecoyTrace scenarios aggregate over a different number of honest nodes (ten versus eight). Our working hypothesis, not yet confirmed by a dedicated ablation, is that this is a consequence of clustering eight nodes instead of ten once the attacker has been blocked. Nevertheless, the pattern remains a useful signal in a more limited sense: it is consistent with containment removing the attacker's influence on the aggregation flow, rather than merely suppressing it while leaving residual contamination, although this interpretation should likewise be read as suggestive rather than demonstrated.

The consistency of this recovery constitutes the central empirical claim of this work. It has been verified across five datasets spanning label-space sizes from 10 classes (MNIST, CIFAR-10) to 100 (CIFAR-100). Furthermore, it holds across three topological densities that emphasize different pivot- and shape-attribution mechanisms, as described in Sections~\ref{sec:experimental-topologies} and \ref{sec:methodology-pivot}. A defense validated only under a favorable topology or a simple classification task would leave open the question of whether its effectiveness is merely an artifact of that specific configuration. The factorial design in Section~\ref{sec:experimental-matrix} was constructed precisely to rule out that possibility. The results demonstrate that the recovery pattern holds up under sparse connectivity, where attribution latency is higher. It also holds up under dense connectivity, where the spread of the semantic footprint to honest neighbors is more severe. It covers both the easiest and most difficult classification tasks evaluated.

The resource-cost aspect of the evaluation supports the same conclusion from a different angle. On MNIST, FashionMNIST, and EMNIST, DecoyTrace consistently reduces process CPU usage and communication volume compared to the baseline and attack configurations, once the topology is dense enough that blocking a single neighbor has a proportional effect. For example, on MNIST with \(edge=0.5\), CPU usage drops from \(89.25\%\) to \(28.21\%\), and on FashionMNIST with \(edge=0.2\), data transmission is reduced by more than half. This efficiency gain is a direct structural consequence of the containment protocol described in Section~\ref{sec:methodology-pivot} and not an incidental benefit. Removing the connection to a blocked neighbor eliminates both the computation and the bandwidth previously dedicated to serving it.

The comparison in Section~\ref{sec:sota} identifies a structural gap in the existing literature: proposals that identify malicious clients, such as FLDetector~\cite{ZhangCJG22} and FedGT~\cite{Xhemrishi2023}, and proposals that use cyber deception, such as HoneyFL~\cite{Zheng2025HoneyFL} and HoneyFED~\cite{MasseBTHSH25}, both assume a central server or an equivalent global vantage point, and none of the reviewed works combines identification, deception, and containment within a serverless DFL topology. The results of Section~\ref{sec:results} demonstrate that this combination is not merely a formal novelty but an empirically viable one. \textit{DecoyTrace} achieves recovery comparable to what a server-mediated defense might promise, using only local, peer-to-peer signals, without ever aggregating a global view of the network's state at a trusted coordinator.

Several limitations of the current design and evaluation are worth stating explicitly, since acknowledging them precisely is more useful than leaving them implicit in the results. The evaluation in this work is confined to a single class of attack, targeted semantic label-flipping, described in Section~\ref{sec:experimental-adversary}, and to a single attacker per configuration. Multiple simultaneous colluding attackers, an attacker that adapts its behavior specifically to evade the chaotic bait's statistical signature rather than merely camouflaging its parameter-space update, or an attacker that deviates from the non-participation assumption of Section~\ref{sec:methodology-threat-model} by actively engaging in the reputation protocol, are outside the scope of the threat model evaluated in this work and remain open questions for future evaluation. The recovery achieved by DecoyTrace is also not uniform across datasets, and CIFAR-10 and CIFAR-100 recover less completely than the three grayscale datasets. On CIFAR-10, both DecoyTrace configurations remain roughly \(0.05\) below Baseline at \(edge=0.5\) and \(edge=1.0\), rather than essentially matching it as observed for FashionMNIST; at \(edge=0.2\) the gap is wider still. On CIFAR-100, Global Reset at \(edge=0.2\) actually falls below Baseline, at \(0.036\) against \(0.090\), the only configuration across the entire evaluation in which a DecoyTrace variant underperforms the undefended Baseline. Finally, the resource-efficiency benefit of containment, clear and consistent for the three grayscale datasets, does not hold uniformly for CIFAR-100, where several DecoyTrace rows show higher CPU utilization than Baseline or Attack at the same topology, for instance \(391.65\%\) for No Global Reset against Baseline's \(332.69\%\) at \(edge=1.0\).

\section{Conclusions}
\label{sec:conclusions}

This work introduced \textit{DecoyTrace}, a proactive cyber deception-based defense designed for strict serverless DFL environments. The proposed approach combines active bait generation, behavioral verification, distributed attribution, topology-aware tracing, and containment into a unified defensive lifecycle. Unlike existing approaches that primarily focus on detecting suspicious updates or excluding malicious participants, DecoyTrace addresses the entire process, from attack exposure to source identification and mitigation, while preserving the federation's decentralized nature.

The experimental evaluation across 60 configurations, including five datasets, three network topologies, and four attack/defense scenarios, demonstrates the feasibility of combining Cyber Deception and DFL security. DecoyTrace consistently recovers a substantial fraction of the utility lost under semantic poisoning attacks, closing most of the gap to undefended benign training on MNIST, FashionMNIST, and EMNIST. For instance, it restores FashionMNIST's honest-node F1-score from an Attack floor of \(0.472\) to within \(0.03\) of Baseline at \(edge=0.5\). In several EMNIST and CIFAR-100 configurations, recovery matches or exceeds Baseline once the attacker is contained. On CIFAR-100 at \(edge=1.0\), for example, the recovered F1-score more than triples the undefended Baseline. Recovery is less complete on CIFAR-10. There, residual contamination and the cost of the recovery mechanisms remain visible in the reported gap to Baseline. This shows that containment alone can be insufficient once poisoned information has already propagated through the federation. It also motivates the integration of explicit recovery mechanisms into the defensive cycle. Beyond utility recovery, the evaluation highlights the relevance of active verification and distributed attribution in decentralized environments. DecoyTrace uses behavioral evidence rather than relying solely on parameter-space similarities, which lets it distinguish malicious behavior from effects propagated through honest intermediate participants. The combination of semantic verification and reputation-based evidence guides tracing and containment without requiring a permanent central coordinator.

Future research will focus on extending DecoyTrace towards larger and more dynamic DFL environments, including experiments with hundreds of participants, changing topologies, and intermittent connectivity, and on evaluating it against stronger adversarial models, including colluding attackers and adversaries that actively attempt to infer or avoid the deployed deception strategy. Another promising direction is replacing the current coordinated reset with lighter recovery mechanisms able to remove residual poisoning effects while preserving accumulated knowledge and minimizing disruption to legitimate training, together with a closer study of the trade-off among defensive coverage, communication overhead, and scalability when deploying multiple DecoyNodes or adaptive deployment strategies.

\section{Acknowledgments} 

This work was supported by Seneca Foundation in Region of Murcia (Spain) (22943/\\FPI/25), co-funded by Scorpion Cybertechnologies, and the SHIELD-6G project, funded by EU’s Horizon Europe research and innovation programme under the SNS JU Grant Agreement No. 101291599. Views and opinions expressed are, however those of the author(s) only and do not necessarily reflect those of the EU or the SNS JU. Neither the EU nor the granting authority can be held responsible for them.

\bibliographystyle{elsarticle-num} 
 \bibliography{cas-refs}

%% else use the following coding to input the bibitems directly in the
%% TeX file.

% \begin{thebibliography}{00}

% %% \bibitem{label}
% %% Text of bibliographic item

% \bibitem{}

% \end{thebibliography}
\end{document}